\documentclass[amsmath,amssymb,pra,aps,showpacs,superscriptaddress,twocolumn]{revtex4-2}
\usepackage{amsmath,amsfonts,amssymb,amsthm,graphics,graphicx,epsfig,bbm}
\usepackage[colorlinks=true,citecolor=blue,linkcolor=red,urlcolor=blue]{hyperref}
\usepackage[usenames]{color}
\usepackage{graphicx}
\usepackage{subfigure}
\usepackage{amsmath}
\usepackage{epsfig}
\usepackage{dcolumn}
\usepackage{bm}
\usepackage{color}
\usepackage{epstopdf}
\usepackage{amssymb}
\usepackage{amstext}
\usepackage{latexsym}
\usepackage{hyperref}
\usepackage{amsfonts}
\usepackage{psfrag}
\usepackage{xcolor}
\usepackage[normalem]{ulem}
\usepackage{dsfont}
\usepackage{txfonts}
\usepackage{overpic}
\usepackage{ifthen}

\usepackage[capitalise]{cleveref}

\newcommand{\ket}[1]{\vert #1 \rangle}
\newcommand{\bra}[1]{\langle #1 \vert}

\newcommand{\braket}[2]{\langle #1 \vert #2 \rangle}

\newcommand{\an}[2]{\ifthenelse{\equal{#1}{}}{\ensuremath{\hat{#1}_{#2}}}{\ensuremath{\hat{#1}^{\protect\phantom{\dagger}}_{#2}}}}

\newcommand{\NTT}{NTT Basic Research Laboratories \& Research Center for Theoretical Quantum Physics, 3-1 Morinosato-Wakamiya, Atsugi, Kanagawa, 243-0198, Japan}

\newcommand{\OIST}{Quantum Information Science and Technology Unit, Okinawa Institute of Science and Technology Graduate University, Onna-son, Okinawa 904-0495, Japan}

\newcommand{\SOKENDAI}{School of Multidisciplinary Science, Department of Informatics, SOKENDAI (the Graduate University for Advanced Studies), 2-1-2 Hitotsubashi, Chiyoda-ku, Tokyo 101-8430, Japan}

\newcommand{\NII}{National Institute of Informatics, 2-1-2 Hitotsubashi, Chiyoda-ku, Tokyo 101-8430, Japan}

\begin{document}

\title{Equilibration of Non-interacting Photons and Quantum Signatures of Chaos}

\author{V. M. Bastidas}
\email{victor.bastidas@ntt.com} 
\altaffiliation[]{These authors contributed equally to this work.}
\affiliation{\NTT}
\author{H. L. Nourse}
\altaffiliation[]{These authors contributed equally to this work.}
\affiliation{\OIST}
\author{A. Sakurai}
\affiliation{\OIST}
\author{A. Hayashi}
\affiliation{\OIST} 
\affiliation{\SOKENDAI}
\affiliation{\NII} 
\author{S. Nishio}
\affiliation{\OIST} 
\affiliation{\SOKENDAI}
\affiliation{\NII} 
\author{Kae Nemoto}
\affiliation{\OIST}
\affiliation{\SOKENDAI}
\affiliation{\NII} 
\author{W. J. Munro}
\affiliation{\NTT}
\affiliation{\NII} 

\date{\today}

\begin{abstract}
Equilibration plays a fundamental role in our understanding of statistical mechanics and the long-time dynamics of many-body systems. In quantum systems, the route to equilibration is intimately related to level repulsion and quantum signatures of chaos that are encoded in their unitary evolution. Chaotic quantum systems exhibit the level statistics characteristic of ensembles of random matrices. In this work, we demonstrate that single-particle chaos leads to equilibration of many non-interacting photons. We show that the underlying mechanisms for equilibration are operator spreading and quantum interference.
More specifically, we demonstrate that the unitary dynamics of a general Floquet system implemented using single-mode phase shifters and multiport beamsplitters leads to equilibration of photons. We propose a realistic photonic implementation of the multiparticle kicked rotor, which is a Floquet system that we use as a concrete example of our general approach.

\end{abstract}

\maketitle

\section{Introduction}

Chaos plays an important role in our daily life and especially in technological applications~\cite{stavroulakis2005chaos,chen2003chaos}. Classically chaotic systems are well known to be extremely sensitive to small perturbations in the parameters that define them~\cite{lorenz1963deterministic,Eckmann1985}. The exploration of these systems is challenging and their importance in our lives is evident from the prediction of weather forecast~\cite{trevisan2011chaos}, the study of turbulence~\cite{chian2022nonlinear} and fluid dynamics to the behavior of financial markets~\cite{chen2008nonlinear}. Of course, such behavior is not restricted to classical systems.

In the quantum world, there are complex systems that exhibit a well-defined semiclassical limit that is chaotic in nature~\cite{Stockmann1990,haake2010,stechel1984quantum}. The investigation of the properties of these quantum systems is not simple because far away from the semiclassical limit the notion of phase-space trajectories is not well defined and one needs to look for quantum manifestations of chaotic behavior~\cite{Breslin1997,haake2010}. These manifestations are referred to as quantum signatures of chaos (QSOC)~\cite{jalabert1994universal,Breslin1997,haake2010,hamma2021a,hamma2021b} and currently there is a plethora of them, ranging from level statistics~\cite{stechel1984quantum,Delande1986,Hentschel2002}, Lochschmidt echoes~\cite{Gutierrez2009,Yan2020}, out-of-time order correlators~\cite{yoshida2017a,yoshida2017b,hamma2021a,Rammensee2018}, and information scrambling~\cite{Shen2020,mi2021information,sahu2022quantum,Harris2022}. In the context of level statistics, it is conjectured~\cite{schmit1984,berry1985,sieber2002,altland2004} that the spectral properties of a system with a chaotic semiclassical limit are related to random matrix theory (RMT) and the symmetries of the system~\cite{Beenakker1997,guhr1998random,Mitchell2010, salazar2024matrixensemblearbitrarycomplex}. Experimental demonstrations of QSOC are abundant in diverse communities such as nuclear physics\cite{guhr1998random}, cold atoms~\cite{Moore1994,Moore1995}, trapped ions~\cite{landsman2019verified} and superconducting qubits~\cite{Roushan2017,Zha2020}.

In statistical physics, it is well known that there are two relevant time scales~\cite{feynman2018statistical}. At short-time scales, fast processes dominate the dynamics far from equilibrium. At long times, after the effect of fast processes has ended, the system reaches a steady state~\cite{feynman2018statistical}. This process is known as equilibration~\cite{Gogolin2016,Wilming2019}. In quantum systems, equilibration has a deep connection to level repulsion~\cite{GarciaMata2015}. When the system has many symmetries and exhibits level clustering, it cannot reach equilibrium~\cite{kinoshita2006quantum}. On the other hand, a chaotic system exhibits strong level repulsion and after time average, the system reaches a steady state characterized by the diagonal ensemble~\cite{Gogolin2016,Wilming2019}. This suggests that there may be an intriguing link between systems that exhibit QSOC and equilibration.

One way to take a system out of equilibrium is to consider the effect of a time-periodic drive. In this scenario, the dynamics can be described using Floquet theory~\cite{bukov2015universal}. In contrast to undriven systems, a periodic drive breaks energy conservation. Furthermore, QSOC in Floquet systems are related to the eigenphases of the unitary operator in one period of the drive~\cite{PilatowskyN2024}. Floquet systems have been shown to exhibit unique behaviors with no counterpart in undriven systems~\cite{bukov2015universal}. Previous works have explored equilibration of driven fermionic systems~\cite{moessner2017equilibration, Lazarides2024}. The nature of many-particle states in bosonic systems such as non-interacting photons drastically deviates from the fermionic states. Currently, a clear understanding of equilibration of driven bosonic systems remains unexplored. It is then natural to ask whether QSOC of a Floquet system at the single-particle level enables a non-interacting photons to equilibrate.

In this work, we establish the relation between QSOC and equilibration of non-interacting photons, i.e., governed by quadratic bosonic Hamiltonians. We discuss QSOC for general photonic systems of non-interacting photons such as level statistics, spectral form factors (SFF), and localization properties of Floquet states. With these results at hand, we define a non-universal photonic out-of-time-order correlator (OTOC) and show that it is related to the permanent of a submatrix of the single-particle unitary evolution operator. We explore how the dynamics are intimately related to the crossover from regular to chaotic behavior at the single-particle level. To substantiate our results, we propose a photonic implementation of the kicked rotor, a paradigmatic model in the community of quantum chaos~\cite{Altland1996,Ammann1988,dArcy2001,Iomin2002,Manai2015,Akridas-Morel2019,santhanam2022quantum}.
Our proposed photonic system is given as a product of phase shifters and a multiport beam splitter~\cite{Kok2007}. We also consider the effect of disorder in the phase shifters. Both the strengths of the disorder and the multiport beam splitters control the crossover from regular to chaotic behavior, allowing the exploration of a wide parameter space. In our work, we focus on photons in optical arrays as their interactions are negligible~\cite{Kok2007}. This is in stark contrast to photons in the microwave regime, which can interact strongly as in the case of superconducting arrays~\cite{kjaergaard2020superconducting}. In cold atoms, one can also investigate the dynamics of multiple bosonic particles. But usually the interactions between them are also strong\cite{schafer2020tools}.

\begin{figure}
	\includegraphics[width=0.40\textwidth]{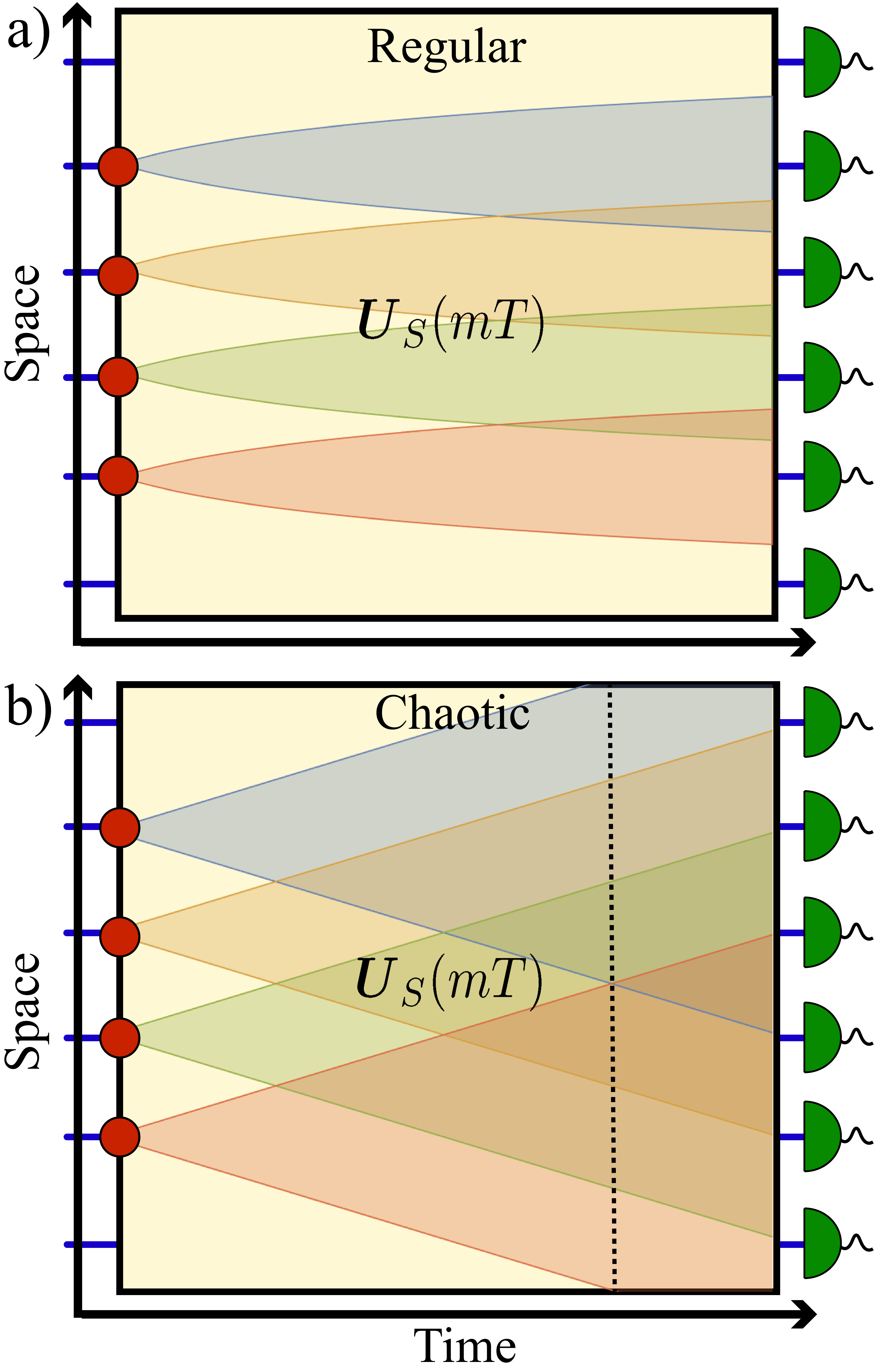}
	\caption{Schematic illustration of regular vs chaotic Floquet dynamics dynamics of $N$ photons in $M$ modes. The dynamics of the modes are generated by an $M\times M$ unitary matrix  $\boldsymbol{U}_{S}(mT)$, where $M$ is the number of modes and $T$ is the period. a) When the dynamics are regular, the photons remain localized with restricted operator spreading (shaded areas) and not all of them are able to interfere.
 b) In the chaotic regime, the operators spread with a typical linear light cone (shaded areas). This allows all the photons to effectively interfere after a characteristic time (dashed line). 
 Due to causality, identical photons can only interfere when their lightcones overlap. This interefence is the underlying mechanism that allows equilibration of local observables at long times.
 }
	\label{Fig0}
\end{figure}

Now in \cref{Fig0} we show a schematic that illustrates the main idea of our work. The intimate relation between QSOC and random matrix theory (RMT) tells us that chaotic systems show universality~\cite{haake2010,Stockmann1990}, and are described by ensembles of random matrices such as the Gaussian orthogonal ensemble (GOE) or the circular orthogonal ensemble COE~\cite{guhr1998random}. We show that the ability of the system to explore the available configurations over time might be related equilibration. When the system is in the chaotic regime, it can explore most of the available configurations. In contrast, in the regular regime it can only access a few of them.

The structure of our paper is as follows. In \cref{SecII} we provide a brief summary of the basic aspects of single-particle Floquet theory and the dynamics of multiple photons. Then in \cref{SecIII} we introduce QSOC, such as quasienergy level statistics, spectral form factors, and localization properties of Floquet states. Next in \cref{SecIV} we show the relation between non-universal photonic OTOCs and the permanent. In \cref{SecV} we discuss how QSOC influence the dynamics of local observables and, in particular, we discuss the relation to equilibration. 
The results presented in the previous sections are general. For this reason in \cref{SecVII}, we provide a specific example of a photonic Floquet system exhibiting QSOC, that is intimately related to the kicked rotor. For this particular example, in \cref{SecVIII} we present numerical results for QSOC, dynamics of observables measurable in experiments, and the statistics of submatrices.
Lastly, we make concluding remarks and an outlook in \cref{SecIX}.

\section{Periodic photonic circuits: Single-particle Floquet theory and the dynamics of multiple photons \label{SecII}}
In our work we will establish a general framework that relates photonic dynamics and QSOC to equilibration. We will extensively use tools of periodically-driven systems theory~\cite{grifoni1998driven}. Let us start by examining the photonic dynamics in the context of Floquet theory for time-periodic Hamiltonians $\hat{H}(t+T)=\hat{H}(t)$ with a period $T$~\cite{grifoni1998driven}. Due to the time-periodicity, it is convenient to define the Floquet operator $\hat{\mathcal{F}}=\hat{U}(T)$ that generates the evolution of the system $\ket{\Psi(mT)}=\hat{\mathcal{F}}^m\ket{\Psi(0)}$ at stroboscopic times $t_m=mT$. 

The advantage of using the language of Floquet theory is that we can use many results of periodically driven quantum systems to understand the properties of the photonic system. In an actual photonic implementation, the coaxial propagation coordinate $z$ of light along an optical waveguide acts as a ``time"~\cite{rechtsman2013photonic}.

Next, let us explore to which extent single-particle Floquet theory relates to the dynamics of multiple photons. We will consider the initial state
 \begin{align}
 \label{eq:InitialState}
\ket{\Psi(0)} & = \ket{n^{(I)}_1,n^{(I)}_2,\ldots,n^{(I)}_M}
=\prod^M_{j=1} \frac{(\hat{a}^{\dagger}_{j})^{n^{(I)}_j}}{\left[n^{(I)}_j!\right]^{1/2}}\ket{\boldsymbol{0}}
  \ ,
 \end{align}
where $\ket{\boldsymbol{0}}=\ket{0_1,0_2,\dots,0_M}$ represents the $M$-mode vacuum.
\cref{eq:InitialState} describes an initial configuration of $N=\sum_j n^{(I)}_j$ particles distributed among $M$ modes, which is labelled by $I$. $n^{(I)}_j$ denotes the number of photons in mode $j$. For simplicity we restrict ourselves to an initial configuration where at most one particle can occupy a given mode ($n^{(I)}_j \in \{0,1\}$). 
We denote the single-particle basis as
\begin{equation} \label{eq:single-particle-basis}
\ket{j} \equiv \hat{a}_j^{\dagger} \ket{\boldsymbol{0}}=\ket{0_1,0_2,\dots,1_j,\dots,0_M},
\end{equation}
such that the $M \times M$ matrix representation of $\hat{\mathcal{F}}^m$, in the single-particle subspace, is defined as
\begin{equation}
[ \boldsymbol{U}_S(mT) ]_{ij} \equiv U_{ij}(mT) \equiv \bra{i} \hat{\mathcal{F}}^m \ket{j}.
\end{equation}
The evolution of the bosonic operators can be expressed in the Heisenberg picture as~\cite{Kok2007,gard2015introduction}
 \begin{equation}
 \label{eq:OperatorEvolution}
\hat{a}^{\dagger}_{i}(mT)=\sum^M_{j=1}U_{i,j}(mT)\hat{a}^{\dagger}_{j}.
 \end{equation}
The typical evolution of the operators is illustrated in \cref{Fig0}, where we can interpret the stroboscopic evolution as a photonic quantum circuit with depth $mT$.

\subsection{Photonic dynamics and permanents}
 
A crucial aspect of the system is that the dimension of the problem's Hilbert space grows with the number of bosons $N$ in a photonic circuit with $M$ modes. Due to the statistics of the photons, the dimension of the Hilbert space is given by~\cite{Feng2019}
\begin{align}
\label{eq:DimensionHilbert}
    ^{M+N-1}C_N=\frac{(M+N-1)!}{N!(M-1)!}
\ ,
\end{align}
which corresponds to the number of $N$-combinations of a set of $M+N-1$ elements~\cite{Barghathi2020}. The number of configurations quickly increases with the number of photons and modes. For example, with $N=3$ photons distributed among $M=12$ modes there are $^{12}C_3=364$ configurations.

The evolution of the quantum state defined in \cref{eq:InitialState} after $m$ periods of the drive is given by
 \begin{align}
 \label{eq:QuantumStrobDyn}
\ket{\Psi(mT)}&=\prod^M_{i=1} \frac{[\hat{a}^{\dagger}_{i}(mT)]^{n^{(I)}_i}}{\left[n^{(I)}_i!\right]^{1/2}}
\ket{\boldsymbol{0}}
\nonumber 
\\
&=\sum_{F}\gamma_{F}(mT)\ket{n^{(F)}_1,n^{(F)}_2,\ldots,n^{(F)}_M}
 \ ,
 \end{align}
 where $F$ denotes all the possible configurations of $N$ bosonic particles in $M$ modes, while $n^{(F)}_j$ is the population of the $j$th mode for a given configuration $F$ such that $N=\sum_j n^{(F)}_j$. The probability amplitude $\gamma_{F}(mT)$ in \cref{eq:QuantumStrobDyn} also defines the matrix elements of the evolution  operator in the $N$-particle subspace~\cite{aaronson2011computational,gard2015introduction}
\begin{align}
\label{eq:MParticleFloquet}
\gamma_{F}(mT)&=\bra{n^{(I)}_1,n^{(I)}_2,\ldots,n^{(I)}_M}\hat{U}(mT)\ket{n^{(F)}_1,n^{(F)}_2,\ldots,n^{(F)}_M}
\nonumber\\
&=\frac{\text{Per}\left[U^{(F,\ I)}(mT)\right]}{\sqrt{n^{(F)}_1!n^{(F)}_2!\cdots n^{(F)}_M!}}
\ . 
\end{align}
This is given in terms of the permanent of an $N\times N$ submatrix $U^{(F,I)}(mT)$ of $\boldsymbol{U}_{\text{S}}(mT)$.
The submatrix $U^{(F,\ I)}(mT)$ depends on the initial configuration $I$ and the specific configuration $F$ measured at the end of the experiment. More specifically, $U^{(F,\ I)}(mT)$ is obtained by keeping $n^{(F)}_j$ copies of the $j$th column and $n^{(I)}_i$ copies of the $i$th rows. Due to the simplification of the initial configuration $I$, we only need to choose one copy of a given row of $\boldsymbol{U}_{\text{S}}(mT)$. 
The corresponding probability of obtaining the configuration $F$ is
\begin{align}
\label{eq:PermanentOutput}
P_{F}(mT)=|\gamma_{F}(mT)|^2=\frac{\left|\text{Per}\left[U^{(F,\ I)}(mT)\right]\right|^2}{n^{(F)}_1!n^{(F)}_2!\cdots n^{(F)}_M!}
\ . 
\end{align}

It is now illustrative to consider the dynamics of the local mean number of photons in mode $l$, given by
 \begin{align}
 \label{eq:LocalMeanPhoton}
 \langle\hat{n}_l(mT)\rangle&= \bra{\Psi(mT)}\hat{a}^{\dagger}_{l}\hat{a}_{l}\ket{\Psi(mT)}
 \nonumber\\
 &=\sum_{F} n^{(F)}_l P_{F}(mT)
 \ .
 \end{align}
We see that measurements with single-photon detectors samples the probability distribution $P_F(mT)$. 
As we will discuss in the following sections, the dynamics in a system that exhibits QSOC are encoded in $\boldsymbol{U}_{\text{S}}(mT)$.

\section{QSOC in photonic Floquet systems\label{SecIII}}
\label{sec:quantum-chaos}
In this section we discuss several QSOC that are of interest in the context of our photonic system and we analyze their dynamical consequences for equilibration. We consider an ensemble of unitary operators, $\mathcal{E}_{U} \equiv \{\hat{U}_w(T)\}$, indexed by $w$. 
In order to investigat QSOC, we examine the properties of the quasienergies, $\{\xi_{\alpha}^w\}$, and single-particle Floquet states, $\{\ket{\alpha^w}\}$, of the Floquet operator, $\hat{U}_w(T)$, defined from the eigenvalue problem $\hat{U}_w(T)\ket{\alpha^w}=\exp{(-\mathrm{i}\xi_{\alpha}^wT/\hbar)}\ket{\alpha^w}$, where $-\hbar\pi/T<\xi^w_{\alpha}<\hbar\pi/T$~\cite{bukov2015universal}. We will compare the spectral statistics of the Floquet operator with ensembles of random matrices in RMT.

\subsection{Quasienergy level statistics} \label{sec:level-statistics}

A standard quantity to distinguish ensembles of random matrices is the probability distribution, $P(r)$, of consecutive level spacing ratios \cite{huse2007,roux2013}
\begin{equation} \label{eq:ratios}
r_{\alpha}^w = \frac{\min(s_{\alpha}^w, s_{\alpha-1}^w)}{\max(s_{\alpha}^w, s_{\alpha-1}^w)},
\end{equation}
where $s_{\alpha}^w = \xi_{\alpha+1}^w - \xi_{\alpha}^w \ge 0$ for adjacent quasienergies $\xi_{\alpha}^w$, ordered by increasing energy. We denote $r$ as the level spacing ratio averaged over the ensemble. \cref{eq:ratios} has been used successfully to investigate quantum signatures of single- and multi-particle chaos~\cite{haake2010,guhr1998random}, eigenstate thermalization hypothesis~\cite{Srednicki1994,borgonovi2016quantum,Murthy2019}, and manybody localization\cite{Lazarides2015,Serbyn2016,Sierant2019}. To compare our system to ensembles of random matrices we compute the spectral average of \cref{eq:ratios}, given by
\begin{equation} \label{eq:average-ratio}
\langle r \rangle = \frac{1}{|\mathcal{E}_U|} \sum_{w} \frac{1}{M-2} \sum_{\alpha} r_{\alpha} ^w,
\end{equation}
where we have averaged over the ensemble. When calculated from the Wigner surmise in RMT $\langle r \rangle_{\mathrm{Poisson}} \approx 0.38629$ (regular regime) and $\langle r \rangle_{\mathrm{GOE}} \approx 0.53590$ (chaotic regime) \cite{roux2013}. Hence, \cref{eq:average-ratio} gives an indication when the unitary ensemble has chaotic dynamics.

Even though the probability distribution $P(r)$ is a standard probe used in systems that exhibit QSOC, it only captures local spectral correlations. It misses important long-range spectral correlations~\cite{hamma2021a,hamma2021b}. Therefore, it is also useful to look at other quantities associated to the ensemble $\mathcal{E}_U$ in order to probe QSOC. From now on, to simplify the notation, we will neglect the index $w$ denoting a given unitary in the ensemble and present ensemble averaged quantities.

\subsection{Spectral form factors} \label{sec:sff}
In the theory of QSOC~\cite{haake2010}, one is often interested in the correlations between quasienergy levels. This is obtained by the spectral form factors (SFF), which are intimately related to scrambling~\cite{Shen2020,mi2021information,sahu2022quantum,Harris2022} and to other QSOC~\cite{haake2010}. The infinite temperature $2N$-point SFF is given by
\begin{align}
\label{eq:2NSpectralFormFactor}
\mathcal{R}_{2N}(mT)=\sum_{\boldsymbol{\zeta},\boldsymbol{\eta}}e^{-\mathrm{i}(\sum_{\zeta\in\boldsymbol{\zeta}}\xi_{\zeta}-\sum_{\eta\in\boldsymbol{\eta}}\xi_{\eta})mT/\hbar}
\ ,
\end{align}
where $\boldsymbol{\zeta}=(\zeta_1,\zeta_2,\dots,\zeta_N)$ and $\boldsymbol{\eta}=(\eta_1,\eta_2,\dots,\eta_N)$. Specifically, we will be interested in the two- and four-point SFF
 \begin{align}
\label{eq:2SpectralFormFactor}
\mathcal{R}_2(mT) & = \sum_{\alpha,\beta}e^{-\mathrm{i}(\xi_{\alpha}-\xi_{\beta})mT/\hbar},
\\
\label{eq:4SpectralFormFactor}
\mathcal{R}_4(mT) & = \sum_{\alpha,\lambda,\beta,\rho}e^{-\mathrm{i}(\xi_{\alpha}+\xi_{\lambda}-\xi_{\beta}-\xi_{\rho})mT/\hbar} .
\end{align}

The SFF exhibits universal features found in chaotic systems, such as a dip, ramp, and plateau, which will be determined by the symmetries of the Floquet operator. These features are governed by correlations in the quasienergy levels with gaps $\Delta_{\alpha \beta}  \equiv \xi_{\alpha} - \xi_{\beta} = \sum_{\lambda = \beta}^{\alpha-1} s_{\lambda}$, that define characteristic time scales $\tau_{\alpha,\beta}=\hbar/\Delta_{\alpha,\beta}$ in terms of the nearest-neighbor level spacings $s_{\lambda}$. Hence, as time increases, the SFF probes quasienergy correlations that are closer and closer together, until it is dominated by the smallest (largest) energy (time) scale. Therefore, to investigate the manifestations of universal features found in chaotic systems, it is also important to study the time evolution of the spectral correlations found in the SFF.

An important time scale is the Heisenberg time, $\tau_H$, associated to the energy gap $\Delta_{\alpha, \alpha+1} = s_{\alpha}$ between adjacent quasienergy levels. It is the smallest energy scale (largest time scale), and hence it is dominated by level repulsion in chaotic systems. It can be estimated as $\tau_H = 2\pi \hbar / \langle s \rangle$ \cite{haake2010}, where $\langle s \rangle$ is the ensemble averaged level spacing. $\tau_H$ is associated with the timescale in finite systems that the discrete energy spectrum can be resolved, and is proportional to the dimension of the Hilbert space, $\tau_H \propto M$. 
The Heisenberg time is captured in the SFF, where it determines the onset of the plateau where the SFF approaches its asymptotic value.

In RMT the two-point SFF for the GOE is characterized by the Heisenberg time, and is given by ~\cite{haake2010,zoller2022}
\begin{align} \label{eq:sff-goe}
\mathcal{R}_2^{\mathrm{GOE}} & (mT) = M^2 r(mT)^2 \nonumber \\
    & + M
    \begin{cases} 
    \frac{mT}{\tau_H} - \frac{mT}{\tau_H} \log \left( 1 + 2 \frac{mT}{\tau_H} \right) & \textrm{for } 0 < mT \le \tau_H, \\
    2 - \frac{mT}{\tau_H} \log \left( \frac{2mT/\tau_H + 1}{2mT/\tau_H - 1} \right) & \textrm{for } mT > \tau_H
    \ ,
    \end{cases}
\end{align}
where $r(mT) = \tau_H \mathcal{J}_1(4MmT/\tau_H)/(2MmT)$, and $\mathcal{J}_1(z)$ is the Bessel function of the first kind~\cite{gradshteyn2014table}. At the Heisenberg time, $\tau_H$, the two-point SFF for the GOE approaches the asymptotic value $\mathcal{R}_2^{\text{GOE}}(t \ge \tau_H) = M$. 

\subsection{Level repulsion and localization properties of Floquet states} \label{sec:level-repulsion}

Whether the system is localized or in the quantum chaotic regime will have profound consequences for equilibration because it will determine the multi-particle interference. Let us first investigate the effect of localization and delocalization on the matrix elements of the Floquet operator in the single-particle basis, under the assumption its spectral statistics is that of RMT.

The Floquet operator can be written in the basis of the single-particle Floquet states as $\hat{\mathcal{F}}=\sum_{\alpha}\ket{\alpha}\bra{\alpha}e^{-\mathrm{i}\xi_{\alpha}T/\hbar}$. Hence, the matrix elements in the single-particle position basis [see \cref{eq:single-particle-basis}] can be written as
 \begin{equation}
 \label{eq:FloquetMatrixElements}
 U_{i,j}(mT)=\bra{i}\hat{U(mT)}\ket{j}=\sum_{\alpha}e^{-\mathrm{i}\xi_{\alpha}mT/\hbar}c_{i,\alpha}c^*_{j,\alpha}
\ ,
 \end{equation}
where $c_{i,\alpha}=\braket{i}{\alpha}$. 

When a Floquet state, $\ket{\alpha}$, is localized in real space it follows that $|c_{i,\alpha}|^2 \approx \exp{(-|i-l_\alpha|/\Lambda_\alpha)}$, where $\Lambda_\alpha$ is the localization length, and $l_\alpha$ is the center of mass of the wavepacket. That is, most of the contribution to the dynamics is from the diagonal matrix elements of $U_{i,j}(mT)$ within a band $|i-j|<\text{Max}(\Lambda_\alpha)$, which is determined by the longest localization length scale. We refer the interested reader to Ref.~\cite{Burrel2007} for more information. In terms of the spectrum, localized Floquet states are related to clustering of levels with very small quasienergy gaps~\cite{haake2010}, and the statistics of the gaps follows a Poissonian distribution because the levels are uncorrelated. Hence, when a photon remains localized in space, it cannot interfere with other photons in remote regions at a distance greater than $\Lambda_{\alpha}$, and there is not enough operator spreading (see next section) for the system to equilibrate. 

Next, let us briefly discuss the onset of thermalization in our Floquet bosonic system and how it affects localization properties of Floquet states. As the Floquet operator is unitary, it can be written in terms of an effective Hamiltonian $\hat{H}_{\text{eff}}$ as $\hat{\mathcal{F}}=e^{-\mathrm{i}\hat{H}_{\text{eff}}T/\hbar}$. To talk about thermalization, it is important to identify the conserved quantities of our Floquet system. Previous works define these conserved quantities for fermionic quadratic Hamiltonians~\cite{Ishii2018}. Here we extend the theory for bosons by defining the operator $b^{\dagger}_{\alpha} \equiv \sum_{r}c^*_{r,\alpha}\hat{a}^{\dagger}_r$ that creates bosonic particles in the $\alpha$ quasienergy state, i.e., $\ket{\alpha} \equiv b^{\dagger}_{\alpha}\ket{\boldsymbol{0}}$. As our system is quadratic in the bosonic operators, the effective Hamiltonian is also quadratic and can be written as a system of free bosons $\hat{H}_{\text{eff}}=\sum_{\alpha}\xi_{\alpha}b^{\dagger}_{\alpha}b_{\alpha}$. This naturally defines a set $\{\hat{\mathcal{I}}_{\alpha}\}$ of conserved quantities $\hat{\mathcal{I}}_{\alpha}=b^{\dagger}_{\alpha}b_{\alpha}$.

In the long-time limit, the system is known to reach a steady state known as the Floquet generalized Gibbs ensemble~\cite{Ishii2018}, described by a density matrix
\begin{align}
\label{eq:FloquetGibbs}
    \hat{\rho}_{\text{GGE}}= \frac{1}{Z} e^{-\sum_{\mu}\Gamma_{\mu}b^{\dagger}_{\mu}b_{\mu}}= \frac{1}{Z} \sum_{\alpha}\ket{\alpha}\bra{\alpha}e^{-\Gamma_{\alpha}\xi_{\alpha}}
    ,
\end{align}
where $\Gamma_{\alpha}=1/k_B T_{\alpha}$ and $T_{\alpha}$ are effective temperatures determined by the conserved quantities, while $Z$ is a normalization constant. Further, we can deduce that $|c_{i,\alpha}|^2\approx e^{-\Gamma_{\alpha}\xi_{\alpha}}/Z$ (see \cref{AppendixB})~\cite{Ishii2018}. Thus, the steady state is Gaussian and determined by different effective temperatures. In certain parameter regimes, the system can heat up to infinite temperatures~\cite{DAlessio2014,Haldar2018,Fleckenstein2021}. In this case, one obtains fully delocalized Floquet states with $|c_{i,\alpha}|^2 \approx 1/M$ and the quasienergies exhibit strong level repulsion following COE statistics, while $\{ \xi_{\alpha} \}$  behave like incommensurable phases~\cite{DAlessio2014}. Intuitively, we expect these systems to equilibrate as the photons can explore more modes and interfere.

\section{Non-universal Out-of-time-order correlators (OTOC) and equilibration} \label{SecIV}
One way to capture long-range spectral correlations is to investigate the dynamics of out-of-time-order correlators (OTOCs)~\cite{Rammensee2018,Borgonovi2019}. It is useful to briefly review the main aspects of this theory. Let us start by defining  
\begin{equation}
\label{eq:GenericOTOC}
\mathcal{C}_{\hat{V},\hat{W}}(t)=\langle[\hat{W}(t),\hat{V}]^{\dagger}[\hat{W}(t),\hat{V}]\rangle
\end{equation}
provided that the operators are local and $[\hat{W}(0),\hat{V}]=0$. This quantity measure the commutativity of the operators as a function of time and gives us information about
operator spreading. By rewritting the commutations we obtain 
\begin{equation}
\mathcal{C}_{\hat{V},\hat{W}}(t)=\mathcal{D}_{\hat{V},\hat{W}}(t)+\mathcal{I}_{\hat{V},\hat{W}}(t)-2\text{Re}\{\mathcal{F}_{\hat{V},\hat{W}}(t)\}
\ ,
\end{equation}
where we have defined $\mathcal{D}_{\hat{V},\hat{W}}(t)=\langle\hat{V}^{\dagger}\hat{W}(t)^{\dagger}\hat{W}(t)\hat{V}\rangle$, $\mathcal{I}_{\hat{V},\hat{W}}(t)=\langle\hat{W}^{\dagger}(t)\hat{V}^{\dagger}\hat{V}\hat{W}(t)\rangle$, and $\mathcal{F}_{\hat{V},\hat{W}}(t)=\langle\hat{W}(t)^{\dagger}\hat{V}^{\dagger}\hat{W}(t) \hat{V}\rangle$. Importantly, the term $\mathcal{F}_{\hat{V},\hat{W}}(t)$ exhibit a universal behavior for chaotic systems~\cite{Rammensee2018,Borgonovi2019}.

In this section, we will define a non-universal $2N$-point OTOC and show that it is equal to calculating the permanent. Hence, the properties of our non-universal OTOC give information about operator spreading and how it affects equilibration in our photonic system. It is important to remark that due to its non-universal and system-specific nature, our OTOC does not provide information of the chaos itself. To access that information we require other QSOC such as measured by level statistics and spectral form factors~\cite{haake2010,zoller2022}.

The simple form of the evolution of the modes motivates us to investigate the following two-point OTOC 
\begin{align}
\label{eq:2OTOC}
\mathcal{C}^{(2)}_{i,j}(mT)& = \bra{\boldsymbol{0}}[\hat{a}^{\dagger}_{i}(mT),\hat{a}_{j}]^{\dagger} [\hat{a}^{\dagger}_{i}(mT),\hat{a}_{j}]\ket{\boldsymbol{0}}
\end{align}
by considering the operators $\hat{W}(mT)=\hat{a}^{\dagger}(mT)$ and $\hat{V}=\hat{a}$ in Eq.~\eqref{eq:GenericOTOC}. We emphasize that our OTOC is non-universal because $\mathcal{F}_{\hat{V},\hat{W}}(t)=0$.
Using \cref{eq:OperatorEvolution} this can be simply evaluated, giving
\begin{align}
\mathcal{C}^{(2)}_{i,j}(mT) & = U_{ij}^*(mT) U_{ij}^{}(mT)
\nonumber \\
& = \sum_{\alpha,\beta}c_{i,\alpha}c^*_{j,\alpha}c^*_{i, \beta}c_{j,  \beta}e^{-\mathrm{i}(\xi_{\alpha}-\xi_{\beta})mT/\hbar}
\nonumber \\
& = P_{F}(mT).
\end{align}
We see that the two-point OTOC is calculated from the permanent as in \cref{eq:MParticleFloquet}.

As we previously discussed, when there is strong level repulsion with RMT spectral statistics, one has $|c_{i,\alpha}|^2\approx e^{-\Gamma_{\alpha}\xi_{\alpha}}/Z$~\cite{Ishii2018}. If the systems heats up to infinite temperature, the Floquet states become delocalized. In this situation, using \cref{eq:FloquetMatrixElements}, the two-point OTOC is approximately
\begin{align}
\label{eq:OTOC2SpectralFormFactor}
\mathcal{C}^{(2)}_{i,j}(mT) \approx \frac{\mathcal{R}_2(mT)}{M^2}
\ .
\end{align}
This naturally establishes the relation between our photonic OTOC and the two-point SFF, $\mathcal{R}_2(mT)$, given by \cref{eq:2SpectralFormFactor}.

Next, let us discuss the four-point OTOC, given by
 \begin{align}
  \label{eq:OTOC4SpectralFormFactor}
 \mathcal{C}^{(4)}_{i,j,r,s}(mT)=\bra{\boldsymbol{0}}\hat{C}^{\dagger}_{i,j,r,s}(mT) \hat{C}_{i,j,r,s}(mT)\ket{\boldsymbol{0}}
  \ ,
 \end{align}
 where $\hat{C}_{i,j,r,s}(mT)=[\hat{a}^{\dagger}_{i}(mT)\hat{a}^{\dagger}_{j}(mT), \hat{a}_{r}\hat{a}_{s}]$. 
 To evaluate this, it is enough to look at the expression
 \begin{align}
  \label{eq:OTOC4Trick}
 \hat{C}_{i,j,r,s}(mT)\ket{\boldsymbol{0}}&=\sum_{o,p}U_{i,o}(mT)U_{j,p}(mT)\hat{a}_{r}\hat{a}_{s}\hat{a}^{\dagger}_{o}\hat{a}^{\dagger}_{p}\ket{\boldsymbol{0}}\nonumber\\
 &=[U_{i,r}(mT)U_{j,s}(mT)+U_{i,s}(mT)U_{j,r}(mT)]\ket{\boldsymbol{0}}
  \ .
 \end{align}
 From this we can directly obtain the 4-point OTOC
 \begin{align}
  \label{eq:FinalOTOC4}
 \mathcal{C}^{(4)}_{i,j,r,s}(mT)=P_{F}(mT)=\left|\text{Per}\left[U^{(F,\ I)}(mT)\right]\right|^2
  \ ,
 \end{align}
where  $\text{Per}\left[U^{(F,\ I)}(mT)\right]$ denotes the permanent of a submatrix
\begin{align}
          \label{eq:SubTwoPhPermanentOutput}
          U^{(F,\ I)}=
    \begin{bmatrix}
    U_{i,r}(mT) & U_{i,s}(mT)  \\
    U_{j,r}(mT) & U_{j,s}(mT)  \\
    \end{bmatrix}
\ , 
\end{align}
of $\boldsymbol{U}_{\text{S}}(mT)$ provided that we initially prepare a two-photon state $\ket{\Psi(0)}=\hat{a}^{\dagger}_{i}\hat{a}^{\dagger}_{j}\ket{\boldsymbol{0}}$ and let the system evolve $m$ periods. This establishes a relation between  $\mathcal{C}^{(4)}_{i,j,r,s}(mT)$, operator spreading, and equilibration of $N=2$ photons. In a similar fashion to the single-particle case, when the system exhibits level repulsion, the correlator $\mathcal{C}^{(4)}_{i,j,r,s}(mT)$ can be written in terms of the four-point SFF, $\mathcal{R}_4(mT)$, given by \cref{eq:4SpectralFormFactor}.

The examples presented so far for few particles gives insight on how to generalize our non-universal  photonic OTOC for multiple particles. In a system with $N$ input photons, the general OTOC corresponds to measuring  a $2N$-point correlator $ \mathcal{C}^{(2N)}_{I,F}(mT)=\bra{\boldsymbol{0}}\hat{C}^{\dagger}_{I,F}(mT) \hat{C}_{I,F}(mT)\ket{\boldsymbol{0}}$, where we define the operator $\hat{C}_{I,F}(mT)=[\prod_{i\in I}[\hat{a}^{\dagger}_{i}(mT)]^{n^{(I)}_i} ,\prod_{j\in F}[\hat{a}_{j}]^{n^{(F)}_j}]$, and $I$ and $F$ are, respectively, the initial and final configurations of $N$ photons in $M$ modes. 
Consequently, the $2N$-point OTOC is given by
\begin{align}
\label{eq:FinalOTOCN}
\mathcal{C}^{(2N)}_{I,F}(mT)=P_{F}(mT)=\frac{\left|\text{Per}\left[U^{(F,\ I)}(mT)\right]\right|^2}{n^{(F)}_1!n^{(F)}_2!\cdots n^{(F)}_M!}
  \ .
\end{align}
In the chaotic regime the $2N$-point photonic OTOC is proportional to the $2N$-point SFF, $\mathcal{R}_{2N}(mT)$, given by \cref{eq:2NSpectralFormFactor}.

The important message we want to convey is that measuring the probability amplitude, $P_{F}(mT)$, 
is equivalent to measuring a photonic OTOC. We will show, with an example, that when the system exhibits QSOC it equilibrates. 
This is linked with the OTOC and scrambling \cite{Rammensee2018} in the chaotic regime, where in a system with QSOC operators spread across all modes, but instead do not in the regular regime. Thus, whether the system equilibrates depends on the periodic photonic chips capability to scramble information and become delocalized [see \cref{Fig0}].

In the next section we will show how the two- and four-point SFF will naturally appear in the dynamics of expectation values and how they determine long-time properties such as equilibration due to level repulsion.

\section{Floquet theory and quantum dynamics of local observables\label{SecV}} 
In this section we discuss another important result of our work. Here we show the intimate relation between QSOC and the dynamics of the system. In particular, we will explore how single-particle signatures of chaos influence the dynamics of observables in a multi-particle scenario. We will show that when the single-particle unitary matrix $\boldsymbol{U}_{\text{S}}(mT)$ [with elements $U_{i,j}(mT)$] exhibits spectral properties related to RMT, the single-particle dynamics  reaches the periodic steady state $\hat{\rho}_{\text{GGE}}$, given by \cref{eq:FloquetGibbs}, that is diagonal in the basis of Floquet states. This is solely determined by the strong repulsion of quasienergy levels that is characteristic of chaotic systems. For simplicity, we focus on the single- and two-particle case, but the results and conclusions presented here are valid for any number of particles. 
 
\subsection{Single-particle dynamics} 
As a first step, it is useful to discuss the effect of QSOC on the dynamics of local observables at the single-particle level. In particular, we will explore the consequences of quasienergy level repulsion with GOE statistics~\cite{haake2010,guhr1998random}. In \cref{SecVII} of our paper, we will show an example of a unitary that exhibits spectral statistics consistent with the Gaussian Orthogonal Ensemble (GOE)~\cite{guhr1998random}.

Let us investigate what happens with the dynamics of a single particle initialized in the state $\ket{\psi(0)}=\hat{a}^{\dagger}_i\ket{\boldsymbol{0}}$ in the regular and chaotic regimes. After $m$ periods of the circuit, the evolution of the particle is given by 
 \begin{equation}
 \label{eq:FloquetBosonicOperator}
 \ket{\psi(mT)}=\sum_{r}U_{i,r}(mT)\hat{a}^{\dagger}_r\ket{\boldsymbol{0}}=\sum_{\alpha}e^{-\mathrm{i}\xi_{\alpha}mT/\hbar}c_{i,\alpha}b^{\dagger}_{\alpha}\ket{\boldsymbol{0}}
   \ .
 \end{equation}

To investigate long-time properties of the system due to level repulsion, such as equilibration~\cite{Lydzba2022generalized}, it is useful to define the time average of observables. For example, the time-averaged number of photons at a given site $l$ over a total number $\mathcal{M}$ of periods of the drive is given by
 \begin{align}
 \label{eq:FloquetAvLocalMeanPhoton}
 \bar{n}_l&=\frac{1}{\mathcal{M}}\sum^{\mathcal{M}-1}_{m=0}\langle\hat{n}_l(mT)\rangle=\frac{1}{\mathcal{M}}\sum^{\mathcal{M}-1}_{m=0}P_{F}(mT)
 \nonumber\\
& =\frac{1}{\mathcal{M}}\sum^{\mathcal{M}-1}_{m=0}\sum_{\alpha,\beta}e^{-\mathrm{i}(\xi_{\alpha}-\xi_{\beta})mT/\hbar}
c_{i,\alpha}c^*_{i,\beta} c_{l,\alpha} c^*_{l,\beta}
 \ ,
 \end{align}
where we have used \cref{eq:LocalMeanPhoton} that relates $ \hat{n}_l$ and $P_{F}(mT)$.
It is worth noticing that the expression for $\bar{n}_l$ resembles the two-point SFF $\mathcal{R}_2(mT)$ in \cref{eq:2SpectralFormFactor}. The dynamics of observables is determined by the quasienergy gaps and the spectral correlations that are encoded in the SFF.

For example, when the system is in the regular regime, the quasienergy gaps become uncorrelated~\cite{haake2010,Stockmann1990} and usually they are also small thus defining a set of long time scales $\tau_{\alpha,\beta}=\hbar/\Delta_{\alpha,\beta}$ discussed above. Further, when the system is close to a degeneracy point or if it exhibit clustering of levels, then it is not able to equilibrate as the average in \cref{eq:FloquetAvLocalMeanPhoton} contains off-diagonal elements $\bra{\beta}\hat{n}_l\ket{\alpha}$ with $\alpha\neq \beta$. On the other hand, when the system exhibits QSOC, the system not only equilibrates but it also thermalizes in the sense of ETH at time scales $\mathcal{M}T\gg\text{Max}(\tau_{\alpha,\beta})$~\cite{Lydzba2022generalized} with
 \begin{align}
 \label{eq:thermalAv}
 \bar{n}_l 
=\sum_{\alpha}|c_{i,\alpha}|^2|c_{l,\alpha}|^2
 \ ,
 \end{align}
where we have used the fact that $\bra{\alpha}\hat{n}_l\ket{\alpha}=|c_{l,\alpha}|^2$. 
This expression can be obtained by assuming that the eigenphases are incommensurable and thus the only term that contribute after the time average is given by the diagonal matrix elements of the observable in the single-particle Floquet basis. In \cref{AppendixB} we provide a formal derivation of this time average. Further, we can deduce that when the system thermalizes~\cite{Ishii2018,Lydzba2022generalized}, $\bar{n}_l=\text{tr}(\hat{\rho}_{\text{GGE}}\hat{n}_l)$ (see \cref{AppendixB}), where $\hat{\rho}_{\text{GGE}}$ is the Floquet generalized Gibbs states in \cref{eq:FloquetGibbs}. At infinite temperature, the Floquet states are fully delocalized in space and $|c_{i,\alpha}|^2=1/M$. Thus, the average local number of photons scales as $\bar{n}_l\sim1/M$. As a consequence of \cref{eq:FloquetAvLocalMeanPhoton,eq:thermalAv}, we obtain the time averaged probability scaling as $\bar{P}_{F}=1/\mathcal{M}\sum^{\mathcal{M}}_{m=1}P_{F}(mT)\sim1/M$.

\subsection{Two-particle dynamics} 
Now let us consider the two-particle case, which will give us insight on the interplay between single-particle chaos and the bosonic character of the particles. Similarly to the single particle case, we investigate the evolution of an initial two particle state $\ket{\Psi(0)}=\hat{a}^{\dagger}_i\hat{a}^{\dagger}_j\ket{\boldsymbol{0}}$. After $m$ periods, the state evolves to
\begin{align}
\label{eq:EvolutionExample}
\ket{\Psi(mT)}&=\hat{a}^{\dagger}_i(mT)\hat{a}^{\dagger}_j(mT)\ket{\boldsymbol{0}}
\nonumber \\
&=\sum^M_{r,s=1}U_{i,r}(mT)U_{j,s}(mT)\hat{a}^{\dagger}_{r}\hat{a}^{\dagger}_{s}\ket{\boldsymbol{0}}
\ .
\end{align}
Next, let us calculate how the single-particle QSOC influence the long-time behavior of 
\begin{align}
\label{eq:TwoPhotonProbability}
P_{F}(mT)&=
|U_{i,r}(mT)U_{j,s}(mT)+U_{i,s}(mT) U_{j,r}(mT)|^2
\ ,
\end{align}
which is obtained from the probability amplitude in \cref{eq:EvolutionAmplitudeExample}.
To see how the level repulsion affects the dynamics, it is convenient to use \cref{eq:FloquetMatrixElements} to write the probability in terms of Floquet states
 \begin{align}
 \label{eq:TwoPhotonProbabilityFloquet}
        P_{F}(mT)&=
         \sum_{\alpha,\lambda,\beta,\rho}e^{-\mathrm{i}(\xi_{\alpha}+\xi_{\lambda}-\xi_{\beta}-\xi_{\rho})mT/\hbar}\left(W_{i,j,r,s}^{\alpha,\lambda,\beta, \rho}+W_{i,j,s,r}^{\alpha,\lambda,\beta, \rho}\right)
          \nonumber \\
        &+2\text{Re}\left[\sum_{\alpha,\lambda,\beta,\rho}e^{-\mathrm{i}(\xi_{\alpha}+\xi_{\lambda}-\xi_{\beta}-\xi_{\rho})mT/\hbar}S_{i,j,r,s}^{\alpha,\lambda,\beta, \rho}\right]
  \ ,
 \end{align}
 where we have defined
 \begin{align}
 \label{eq:TwoPhotonFloquetTensor}
        W_{i,j,r,s}^{\alpha,\lambda,\beta, \rho}&=c_{i,\alpha}c_{r,\alpha}^*c_{j,\lambda}c_{s,\lambda}^*
        c_{i,\beta}^*c_{r,\beta}c_{j,\rho}^*c_{s,\rho}
                   \nonumber \\
       S_{i,j,r,s}^{\alpha,\lambda,\beta, \rho}&=c_{i,\alpha}c_{r,\alpha}^*c_{j,\lambda}c_{s,\lambda}^*
        c_{i,\beta}^*c_{s,\beta}c_{j,\rho}^*c_{r,\rho}  \ .
 \end{align}
 As we are exploring here the dynamics of two particles, it is expected that the dynamics strongly depends on correlations between four single-particle quasienergy levels, which is captured the spectral form factor $\mathcal{R}_4(mT)$ [see \cref{eq:4SpectralFormFactor}].

 In \cref{AppendixB} we describe in detail how to perform the time average of this quantity in the chaotic regime to obtain
 \begin{align}
 \label{eq:TwoPhotonTensorAverage}
       \bar{P}_{F}&=\sum_{\alpha,\lambda}(W_{i,j,r,s}^{\alpha,\lambda,\alpha,\lambda}+W_{i,j,r,s}^{\alpha,\lambda,\lambda,\alpha}+W_{i,j,s,r}^{\alpha,\lambda,\alpha,\lambda}+W_{i,j,s,r}^{\alpha,\lambda,\lambda,\alpha})
       \nonumber \\
       &+2\text{Re}\left[\sum_{\alpha,\lambda}(S_{i,j,r,s}^{\alpha,\lambda,\alpha,\lambda}+S_{i,j,r,s}^{\alpha,\lambda,\lambda,\alpha})\right]
       \ .
 \end{align}
The most important information we want to extract from this time average is the scaling of the probability $\bar{P}_{F}\sim 1/M^2$ for each configuration because $W_{i,j,r,s}^{\alpha,\lambda,\lambda,\alpha}\approx S_{i,j,r,s}^{\alpha,\lambda,\lambda,\alpha}\approx 1/M^4$ (assuming that the single-particle system thermalizes at infinite temperature). This is the approximate scaling that the probability $\bar{P}_{F}$ reaches when the system equilibrates.

We can also write the expression for the state in terms of Floquet states
 \begin{align}
 \label{eq:TwoParticleFloquet}
 \ket{\Psi(mT)}=\sum_{\alpha,\lambda}e^{-\mathrm{i}(\xi_{\alpha}+\xi_{\lambda})mT/\hbar}c_{i,\alpha}c_{j,\lambda}b^{\dagger}_{\alpha}b^{\dagger}_{\lambda}\ket{\boldsymbol{0}}
 \ .
 \end{align}
As expected, the evolution of two photons is determined by a two-particle quasienergy $E_{\alpha,\lambda}=\epsilon_{\alpha}+\epsilon_{\lambda}$ and is given by a quantum superposition of two particles occupying the different available Floquet states. This equation contains valuable information as the overlaps $c_{i,\alpha}=\braket{i}{\alpha}$ and $c_{j,\lambda}=\braket{j}{\lambda}$ contain information about localization properties of the Floquet states, the two particle quasienergies carry information about spectral properties of the system, and the operators $b^{\dagger}_{\alpha}$ allows us to keep track of the bosonic character of the photons.
We can show that
 \begin{align}
 \label{eq:FloquetTwoParticleAvLocalMeanPhoton}
 \bar{n_l}
=\frac{1}{\mathcal{M}}\sum^{\mathcal{M}}_{m=0}\sum_{\alpha,\lambda,\beta,\rho}e^{-\mathrm{i}(\xi_{\alpha}+\xi_{\lambda}-\xi_{\beta}-\xi_{\rho})mT/\hbar}O_{i,j}^{\alpha,\lambda,\beta, \rho} \bra{\boldsymbol{0}}b_{\beta}b_{\rho}\hat{n}_l b^{\dagger}_{\alpha}b^{\dagger}_{\lambda}\ket{\boldsymbol{0}}
 \ ,
 \end{align}
where $O_{i,j}^{\alpha,\lambda,\beta, \rho}=c_{i,\alpha}c_{j,\lambda}c^*_{i,\beta}c^*_{j,\rho}$.

In contrast to the single-particle case, here we need to be cautious and take into account the bosonic character of the particles. For this reason, let us investigate in detail the vacuum expectation value 
 \begin{align}
 \label{eq:VacuumProductBoson}
 \bra{\boldsymbol{0}}b_{\beta}b_{\rho}\hat{n}_l b^{\dagger}_{\alpha}b^{\dagger}_{\lambda}\ket{\boldsymbol{0}}
 &=\sum_{\sigma,\eta}c_{l,\sigma}c^*_{l,\eta}\bra{\boldsymbol{0}}b_{\beta}b_{\rho}\hat{b}^{\dagger}_{\sigma}\hat{b}_{\eta} b^{\dagger}_{\alpha}b^{\dagger}_{\lambda}\ket{\boldsymbol{0}}
 \nonumber\\
 &=c_{l,\rho}c^*_{l,\lambda}\delta_{\alpha,\beta}+c_{l,\rho}c^*_{l,\alpha}\delta_{\beta,\lambda}
  \nonumber\\
 &+c_{l,\beta}c^*_{l,\lambda}\delta_{\alpha,\rho}+c_{l,\beta}c^*_{l,\alpha}\delta_{\rho,\lambda}
 \ ,
 \end{align}
where we have used that $\hat{n}_l=\hat{a}^{\dagger}_{l}\hat{a}_{l}=\sum_{\sigma,\eta}c_{l,\sigma}c^*_{l,\eta}\hat{b}^{\dagger}_{\sigma}\hat{b}_{\eta}$ and the identity
 \begin{align}
 \label{eq:IdentityVacuumProductBoson}
\bra{\boldsymbol{0}}b_{\beta}b_{\rho}\hat{b}^{\dagger}_{\sigma}\hat{b}_{\eta} b^{\dagger}_{\alpha}b^{\dagger}_{\lambda}\ket{\boldsymbol{0}}
&=\delta_{\rho,\sigma}(\delta_{\eta,\lambda}\delta_{\alpha,\beta}+\delta_{\eta,\alpha}\delta_{\lambda,\beta})
\nonumber\\
 &+\delta_{\beta,\sigma}(\delta_{\eta,\lambda}\delta_{\alpha,\rho}+\delta_{\eta,\alpha}\delta_{\lambda,\rho})
\ .
 \end{align}

In the chaotic regime, there are no degeneracies in the quasienergy spectrum and by using the relations discussed above we can explicitly derive the time average in \cref{eq:FloquetTwoParticleAvLocalMeanPhoton} as
 \begin{align}
 \label{eq:FinalTwoPhotonTimeAverage}
 \bar{n_l}&\approx\sum_{\alpha,\lambda}O_{i,j}^{\alpha,\lambda,\alpha, \lambda} |c_{l,\lambda}|^2+O_{i,j}^{\alpha,\lambda,\lambda, \alpha} |c_{l,\alpha}|^2
 \nonumber\\
 &+\sum_{\alpha,\lambda}O_{i,j}^{\alpha,\lambda,\lambda, \alpha} |c_{l,\lambda}|^2+O_{i,j}^{\alpha,\lambda,\alpha, \lambda} |c_{l,\alpha}|^2
 \ .
 \end{align}
Similarly to the single-particle case,  the average local number of photons scales as $ \bar{n}_l\sim 1/M$ when the system equilibrates. 

At this point, it is important to emphasize that at the single-particle level we assume a general system with RMT level statistics. From the theory of generalized thermalization of Floquet systems,
 this implies that such a system thermalizes at a given temperature determined by conserved quantities as in Ref.~\cite{Lydzba2022generalized}. In the multiparticle case, however, the local observables do not thermalize but they equilibrate~\cite{Lydzba2022generalized}. In fact, in a recent experiment, local equilibration~\cite{Somhorst2023} was observed in an optical simulation of undriven Hamiltonians. The scalings that we have obtained from the time averages give us some intuition of the values of observables after equilibration takes place.

From the results presented in this section, we can see the intimate relation between the calculation of time averages and spectral correlations. We can also see how to generalize this to the $N$-particle case. In this situation, the dynamics is determined by $2N$-point level correlations encoded in the spectral form factor $\mathcal{R}_{2N}(mT)$. The level repulsion set the time scales for equilibration~\cite{Lydzba2022generalized}.

\section{An example Model: A hybrid optical Floquet circuit\label{SecVII}}
The results presented so far are general and valid for any unitary operator. Our aim in this section is to provide a concrete example of a system that undergoes a crossover from regular to chaotic behavior.

Next, we will propose a time-periodic photonic system, where within a period of the drive, $T$, the evolution is given by a succession of two operators. First one applies a pattern of local phase shift unitaries, given by
\begin{align} 
\label{eq:phase-shifter}
\hat{U}_1 \equiv \prod_{j=1}^M \exp\left(-\mathrm{i}  \tilde{\phi}_j \hat{a}_j^{\dagger}\hat{a}_j\right),
\end{align}
where $\hat{a}_j^{\dagger}$ creates a photon in mode $j$, $\hat{a}_j$ annhiliates a photon, $\tilde{\phi}_j$ is the angle of the phase shifter, and $M$ is the total number of modes. Then a unitary is applied that acts like an $M$-port beam splitter characterized by an angle $\theta$, given by
\begin{align} 
\label{eq:m-splitter}
\hat{U}_2 \equiv \exp\left[-\mathrm{i}  \theta \sum^{M}_{j=1} (\hat{a}_j^{\dagger}\hat{a}_{j+1}+\hat{a}_{j+1}^{\dagger}\hat{a}_{j})\right].
\end{align}
Hence, the evolution operator in one period of the drive is given by $\hat{\mathcal{F}}  \equiv \hat{U}_2 \hat{U}_1$, where 
\begin{align}
\label{eq:UnitarySingleParticle}
\hat{\mathcal{F}} = \exp\left[-\mathrm{i}  \theta\sum^{M}_{j=1} (\hat{a}_j^{\dagger}\hat{a}_{j+1}+\hat{a}_{j+1}^{\dagger}\hat{a}_{j})\right]\prod_{j=1}^M \exp\left(-\mathrm{i}  \tilde{\phi}_j\hat{a}_j^{\dagger}\hat{a}_j\right)
\ ,
\end{align}
is the Floquet operator~\cite{grifoni1998driven,haake2010}.

\begin{figure}
	\includegraphics[width=0.42\textwidth]{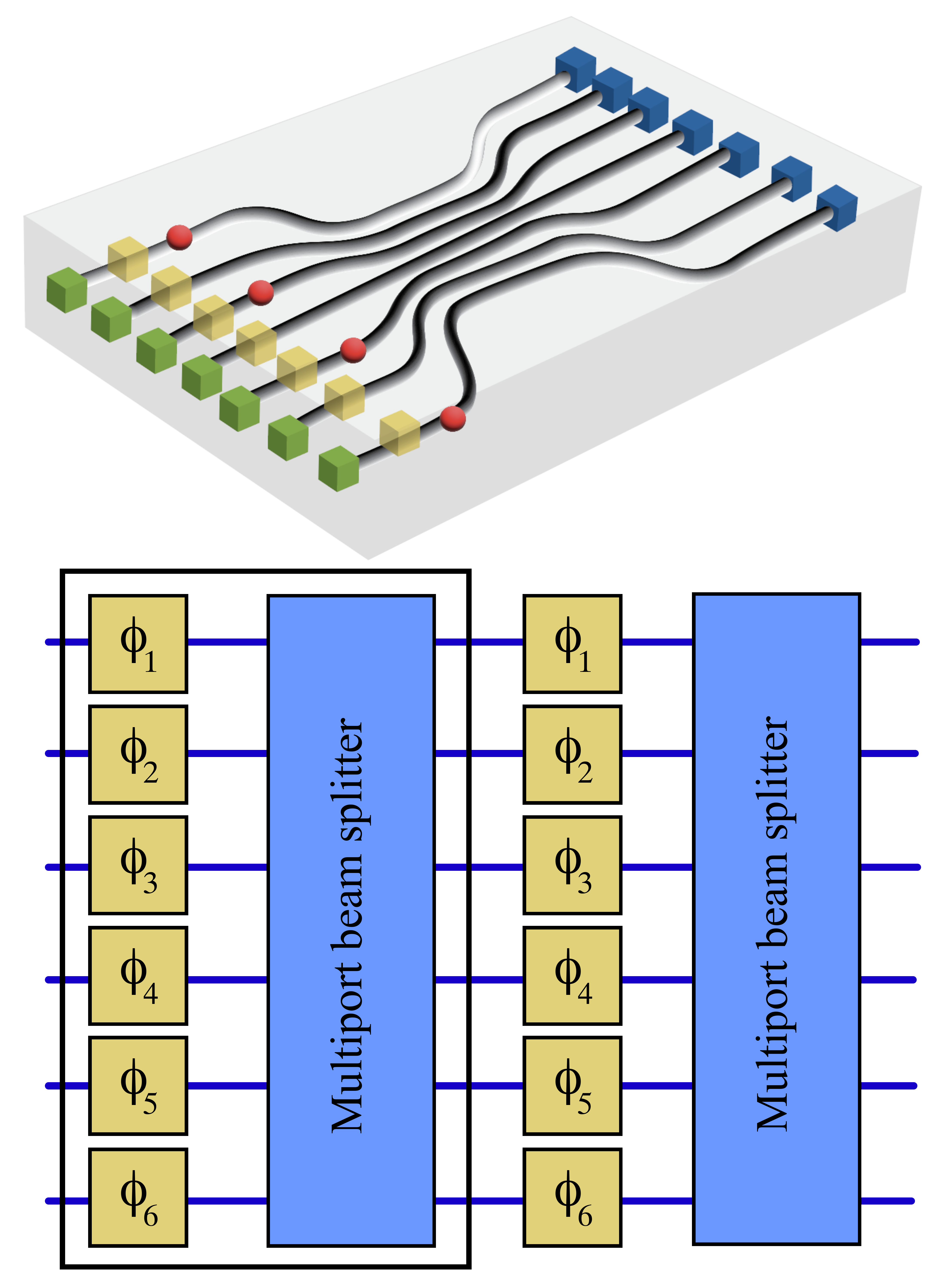}
	\caption{Schematic of a photonic chip that implements the dynamics of the kicked rotor (see \cref{eq:SingleParticle}), for demonstrating equilibration of multiple photons. The yellow boxes represent the local phase-shifter $\hat{U}_1$ (see \cref{eq:m-splitter}). The multiport beam splitter is achieved by bringing the waveguides together, which implements the unitary $\hat{U}_2$ (see \cref{eq:m-splitter}. The black box is one cycle of the drive, whose dynamics is given by the Floquet operator $\hat{\mathcal{F}}$ (see \cref{eq:UnitarySingleParticle}).
 }
	\label{Fig1}
\end{figure}

The unitary given by \cref{eq:UnitarySingleParticle} is general. We propose that it can be physically realized in silica-on-silicon waveguide circuits consisting of $M$ accessible spatial modes~\cite{spring2013boson,gard2015introduction}, which we schematically depict in \cref{Fig1}. For a period of the drive, the waveguides are separated at the beginning to avoid evanescent coupling and phase-shifters are used to implement $\hat{U}_1$. Then, as the photons travel along the chip, the waveguides are quickly brought together, allowing for evanescent coupling~\cite{rechtsman2013photonic}, which implements $\hat{U}_2$. The strength of the evanescent coupling controls the parameter $\theta$.

\subsection{Quantum kicked rotor} 
To obtain some physical intuition of the dynamics generated by our hybrid quantum circuit, it is useful to consider a time-periodic Hamiltonian $\hat{H}(t)=\hat{H}(t+T)$, whose unitary time evolution in one period is given by \cref{eq:UnitarySingleParticle}. The corresponding Hamiltonian is a kicked rotor, given by

\begin{figure*}
	\includegraphics[width=\textwidth]{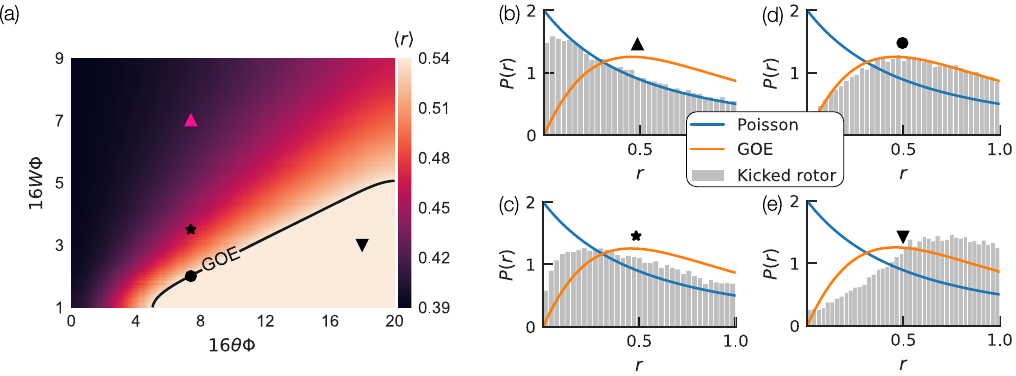} 
	\caption{There is a crossover from Poissonian to Gaussian Orthogonal Ensemble (GOE) statistics in the consecutive level spacing ratio, $r$. (a) The average $\langle r \rangle$, where $\theta$ is the rotation angle of the $M$-port beam splitter, and $\Phi$ is the strength of the harmonic trapping potential. The contour line delineates $\langle r \rangle = 0.53590$ for the GOE. The probability distribution of consecutive level spacing ratios, $P(r)$, is depicted in (b) $W = 7/(16\Phi)$ and $\theta = 7.4/(16\Phi)$ (upward triangle); (c) $W = 3.5/(16\Phi)$ and $\theta = 7.4/(16\Phi)$ (star); (d) $W = 2/(16\Phi)$ and $\theta = 7.4/(16\Phi)$ (circle); (e) $W = 3/(16\Phi)$ and $\theta = 18/(16\Phi)$ (downward triangle). Calculated for $|\mathcal{E}_U| = 100$ disorder realizations, $M=300$ modes, $\Phi = \pi/4$, and $T=1$.
}
	\label{Fig2}
\end{figure*}

\begin{align}
\label{eq:SingleParticle}
\hat{H}(t)=\sum^M_{j=1}\frac{\hbar\tilde{\phi}_j}{T} \hat{a}_j^{\dagger}\hat{a}_j +  \frac{\hbar\theta}{T} \sum^{\infty}_{m=-\infty}\delta(t/T-m) \sum^{M}_{j=1} (\hat{a}_j^{\dagger}\hat{a}_{j+1}+\text{h.c})
\ , 
\end{align}
where the first term is a spatial modulation of onsite energies, and the second term is a periodic train of delta kicks, with kicking strength $\hbar \theta / T$, that couples nearest-neighbor modes.
We define a spatial profile of the onsite angular frequency detunings
\begin{align}
\label{eq:OnSiteFreq-disorder}
\tilde{\phi}_j =\phi_j+\delta_j, 
\end{align}
where
\begin{equation}
\label{eq:OnSiteFreq}
 \phi_j=\frac{4\Phi}{M^2}\left(j-\frac{M}{2}\right)^2          
 \ ,
\end{equation}
acts as a harmonic trapping potential with strength $\Phi$, and $\delta_j$ is a random angle drawn from a uniform distribution in the interval $[-W,W]$.

It is important to emphasize that the Hamiltonian given by \cref{eq:SingleParticle} does not contain interactions between the particles, i.e., it is quadratic in the bosonic operators. As the particles are non-interacting, each particle independently evolves under the time evolution operator $\hat{U}(mT)$. However, interesting physics, such as multiparticle interference~\cite{tichy2012many,Dittel2018}, occurs in the case of multiple bosonic particles due to their statistics~\cite{walschaers2016statistical}.

In our photonic implementation, by adding disorder to the phase-shifters, we define an ensemble of unitaries associated to the dynamics generated by the quantum kicked rotor. The effect of a small amount of disorder is to break any remaining symmetries in our system. At the single particle level this allows the system to thermalize, as has been recently reported in the context of undriven systems~\cite{Lydzba2022generalized}. If the disorder is too strong, however, the system becomes localized. In the next section, we will show that both $\theta$ as well as $W$ are key parameters describing the transition between regular and chaotic behavior in our system.

\subsection{Classical kicked rotor and chaos}
Following the procedure outlined in~\cite{Glaetzle2017}, in the absence of disorder $(W=0)$, one can obtain a semiclassical limit of \cref{eq:SingleParticle} in the single-particle subspace. Here the size $M$ of the chain determines the effective Planck constant given by $\hbar_{\text{eff}}=\hbar/M$. In this way, we can derive a semiclassical Hamiltonian for  $M\gg1$ in terms of the dimensionless canonical variables $x$ and $k$ [see \cref{AppendixA} for its derivation], given by
 \begin{equation}
 \label{eq:MTSinParClassHam}
          \mathcal{H}_{\text{SC}}(x,k,t)=\frac{4\hbar \Phi}{T} \left(x-\frac{1}{2}\right)^2 + \frac{2\hbar\theta}{T} \sum^{\infty}_{m=-\infty}\delta(t/T-m)\cos (k)
  \ .
 \end{equation}
The resulting Hamiltonian has a very similar form to the classical kicked rotor, which is a paradigmatic model of chaotic dynamics~\cite{Altland1996,Ammann1988,dArcy2001,Iomin2002,Manai2015,Akridas-Morel2019,santhanam2022quantum}. In \cref{AppendixA} we show that \cref{eq:MTSinParClassHam} exhibits a crossover from regular to chaos~\cite{santhanam2022quantum}. For example, when $\theta = 1 / (8\Phi)$ the system is regular with a mixed phase space, while for $\theta = 5 / (8\Phi)$ the system is fully chaotic.

The semi-classical limit is obtained in the limit of an infinite chain. However, in experimental platforms one works with a finite number of sites $M$ -- usually a few of them -- far from the semiclassical limit. A natural question is whether the chaotic dynamics found in the classical kicked rotor are also exhibited in the finite sized quantum kicked rotor. We will show in that the quantum kicked rotor defined in \cref{eq:SingleParticle} exhibits a crossover to a regime that exhibits QSOC associated with the classical rotor defined in \cref{eq:MTSinParClassHam}.

\section{Numerical results\label{SecVIII}} 
The purpose of this section is to show numerical results for our particular example of a photonic Floquet circuit. Numerically, we generate $w = 1, \ldots, |\mathcal{E}_U|$ realizations of the Floquet operator \cref{eq:UnitarySingleParticle} which are uniformally distributed with probability $p_w = 1 / |\mathcal{E}_U|$. We show results for the level statistics, spectral form factors and the dynamics of observables. 
\subsection{Quasienergy level statistics}
\cref{Fig2} shows $\langle r \rangle$ [see \cref{eq:average-ratio}] as a function of the $M$-port beam splitter rotation angle (kicking strength), $\theta$, and onsite disorder strength $W$. When disorder dominates, the system is localized in real space with a Poisson distribution. For large kicking that dominates over disorder, pseudoconservation of crystal momentum is recovered and the system is instead localized in momentum space. The competition between kicking and disorder promotes quasienergy level repulsion ($P(r) \sim r$ for GOE), giving large regions where the level statistics appears chaotic.

In the regions where $\langle r \rangle \ne \langle r \rangle_{\mathrm{GOE}}$, the probability distribution $P(r)$ deviates from the ones exactly calculated from the Wigner surmise, as shown in \cref{Fig2}.  In fact, there is a crossover between regions, as is expected for systems exhibiting single-particle chaos~\cite{haake2010}. The quasienergy level statistics gives an indication of where the system is chaotic, and we will confirm this by looking for universal features found in the SFF for systems that exhibit QSOC.

\subsection{Spectral form factors}
\begin{figure}
	\includegraphics[width=\columnwidth]{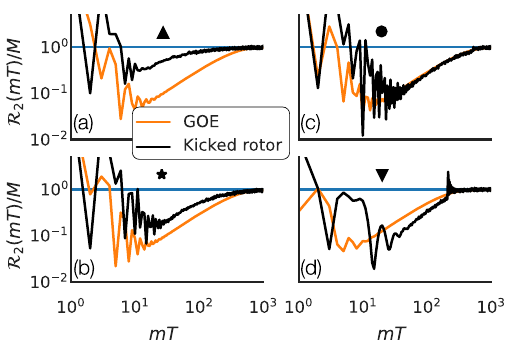}
	\caption{Stroboscopic dynamics of the two-point spectral form factors, $\mathcal{R}(mT)$ [see \cref{eq:2SpectralFormFactor}], shows the characteristic dip, ramp, and plateau of chaotic systems. When the spectral statistics are (a) Poissonian the SFF is close to the long-time asymptote $\bar{\mathcal{R}}_2 = M$ (blue horizontal line). Instead, in (c)-(d) the SFF more closely resemebles that expected in a chaotic system [see \cref{eq:sff-goe}], with a dip, ramp, and plateau. In the crossover regime (b) the kicked rotor exhibits weak QSOC. We set the Heisenberg time as (a) $\tau_H \approx 535T$, (b) $585T$, (c) $605T$, (d) $300T$. The upwards triangle, star, circle, and downwards triangle correspond to the subfigures in \cref{Fig2}. Calculated for $|\mathcal{E}_U|=1000$ disorder realizations and $M=300$ modes.
\label{Fig3}
 }
\end{figure}

For a general chaotic photonic system with $N\ll M$ the SFF $\mathcal{R}_{2N}(mT)$ [see \cref{eq:2NSpectralFormFactor}] shows a typical dynamical behavior characterized by a decay from a value $M^{2N}$ reaching a dip after $m\approx\sqrt{M}$ iterations of the Floquet operator. Finally, the SFF  reaches a plateau at $m \sim M$ with a value $N M^{N}$ that defines the long-time asymptote~\cite{yoshida2017b}. As expected, the SFF reaches a plateau at a time that scales with $M$ as our estimate $\tau_H$. \cref{Fig3} shows the long-time dynamics of the SFF $\mathcal{R}_2(mT)$ [see \cref{eq:2SpectralFormFactor}] in the single-particle sector. 

In the regular regime [see \cref{Fig3}(a)] the SFF displays a dip, but not a pronounced ramp before reaching the plateau. This indicates, along with the quasienergy statistics in \cref{Fig2}(b), that in this region the system is not chaotic. We will later show that in this region the system may not equilibrate. In the chaotic regions [see \cref{Fig3}(c)(d)] the SFF displays the typical features associated with systems that exhibit QSOC. However, even in the crossover region [see \cref{Fig3}(b)] there are weak QSOC, exhibiting a dip, ramp, and plateau. 

\subsection{Long-time dynamics of local observables}
In the this section we will show numerical evidence of the dynamics of $\langle\hat{n}_l(mT)\rangle$ and the probability $P_{\text{F}}(mT)$ for our photonic circuit with $N=2$ photons in $M=12$ modes.  In this case, the single-particle unitary can be represented as a $12\times12$ unitary matrix $\boldsymbol{U}_S(mT)$. This is an interesting example, as the system has a small system size far away of the semiclassical regime $M\gg 1$.

\begin{figure*}
	\includegraphics[width=0.95\textwidth]{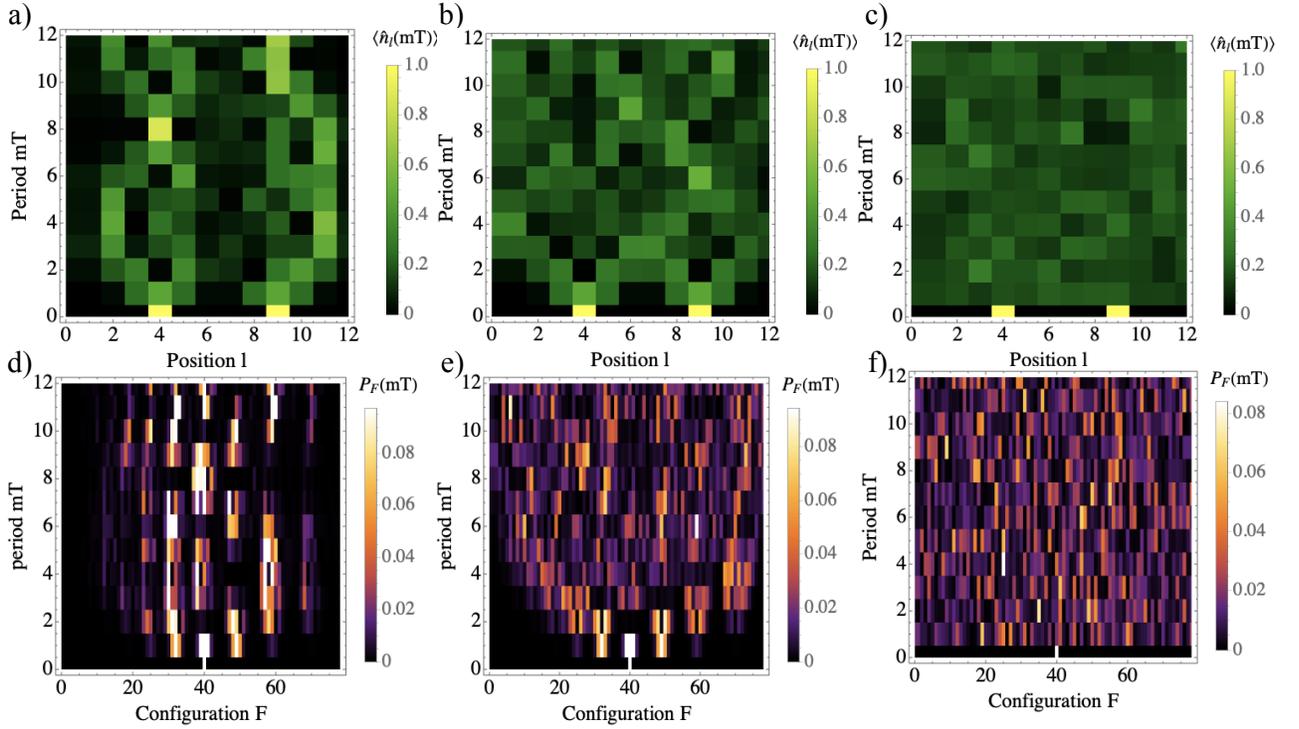}
	\caption{Dynamics over $m=12$ cycles for $N=2$ photons in $M=12$ modes for a single realization of disorder. $a)$ ($W=7/16\Phi$) and $b)$ ($W=1/8\Phi$) depict the dynamics of the mean number of photons $\langle\hat{n}_l(mT)\rangle$ for regular and chaotic unitaries, respectively. $d)$, $e)$ show  $P_{\text{F}}(mT)$ for regular and chaotic unitaries, respectively. Clearly, in the regular regime, the system only explores a small portion of the available configurations. We benchmark our results using the a unitary evolution drawn from the Haar measure in $c)$ and f). The dynamics in $a)$ and $b)$ resemble the light cone structure illustrated in ~\cref{Fig0}. For the simulation we set $\Phi=\pi/4$ and $\theta = 7.4/16 \Theta$.
 }
	\label{Fig4}
\end{figure*}

In \cref{Fig4} we plot the stroboscopic dynamics of  $\langle\hat{n}_l(mT)\rangle$ in the regular regime for $N=2$ photons. From this one can see that the photons remain localized and there are regions of the chain that cannot be accessed as we show in \cref{Fig4}~a). On the contrary, when the system approaches the chaotic regime, it can explore more modes along the lattice, as we show in \cref{Fig4}~b). To benchmark our results, \cref{Fig4}~c) depict the dynamics of the local mean number of photons when the evolution operator is a random matrix drawn from the circular unitary ensemble (CUE)~\cite{guhr1998random,aaronson2011computational}. When the our system approaches the chaotic regime, the dynamics is very close to the case of a random matrix as can be seen in \cref{Fig4}. In these numerical results, we show the dynamics during $m\approx M=12$ periods, which is close to the time scale required to reach the Plateau of the SFF in \cref{Fig3}. 

Next it is illustrative to investigate the populations $P_{F}(mT)$ of the different configurations. The dynamics of multiple photons can be interpreted as a quantum walk in the Hilbert space~\cite{Estarellas2020}. In \cref{Fig4}~(d)~and~(e) we plot the time evolution of the probabilities $P_{F}(mT)=|\gamma_{F}(mT)|^2$ in the regular and chaotic regimes. For comparison \cref{Fig4}~f) shows the dynamics of $P_{F}(mT)$ for a random matrix drawn from CUE.

The results presented so far show that single-particle QSOC lead to equilibration of observables. We also show the intimate relation between spectral statistics, spectral form factors and photonic OTOCs and discuss the role that they play on the dynamics. 

\section{Conclusions\label{SecIX}} 
In summary, we have investigated how QSOC in Floquet systems of non-interacting bosons influence equilibration. We define a non-universal photonic OTOC that provides information of operator spreading for local bosonic operators. In the case of $N$ photons in $M$ modes, our $2N$-point OTOC is exactly the permanent of a $N\times N$ submatrix of the evolution operator. This naturally shows that at short-time scales the system cannot equilibrate because there is not enough operator spreading. The latter is a fundamental ingredient for the system to reach equilibration via quantum interference. Of course, due to their non-universal nature, our $2N$-point OTOC does not provide information of chaos. For this reason, we provide a rigorous analysis of QSOC in our driven system such as level statistics and spectral form factors.

In future works, it would be interesting to explore how to use ideas of condensed matter physics to further break symmetries of the unitary. For example, this may allow one to explore chaotic systems with CUE spectral statistics by breaking time reversal symmetries. In real implementations, the photonic system is affected by photon loss. Therefore, a natural question is to explore how these loss mechanism affects the chaotic dynamics in our photonic implementation. Recently, a duality relating beam splitters to parametric down-conversion processes has been proposed~\cite{cerf2020two,salazar2023linear}. One possible future direction of research is to investigate chaos in dual models with beam splitter and two-mode parametric amplifiers.

Another interesting direction of research is to explore the relation bewteen the dynamics of driven photonic systems and the complexity of Boson Sampling~\cite{aaronson2011computational}. In their original paper~\cite{aaronson2011computational}, Aaronson and Arkhipov show that a unitary from the CUE will be sufficient for boson sampling to be hard. A recent work~\cite{PilatowskyN2023} has shown that the dynamics of Floquet and time-independent systems cannot emulate unitaries drawn from the Haar measure. As a consequence, our Floquet system cannot be used to perform boson sampling. In fact, operator spreading and delocalization of Floquet states may not be enough to ensure that the dynamics of a $N$ photon system has the complexity required for sampling problems. However, Ref.~\cite{PilatowskyN2023} provides an interesting perspective for future directions of research. They discuss a concept called "Complete Hilbert-Space Ergodicity". In a nutshell, they propose the use of quasiperiodic drives to achieve unitaries that are arbitrarily close to the ones drawn from the Haar measure. In fact, quasiperiodic drives are close to any $t$-design for any $t$. It would be interesting to investigate photonic quantum circuits implemmenting a quasiperiodic drive. In this context, it is interesting to investigate which circuit depth is required to generate a desired $t$-design and if such a depth can be achieved in current photonic architectures while preventing significant photon loss.

{\it{Acknowledgments.---} }
V. M. Bastidas thanks M. M. Zapata-Alvarez and valuable discussions with K. Azuma, A. Harrow, H. W. Lau, L. Ruks, and H. Takesue. The authors acknowledge partial support through the MEXT Quantum Leap Flagship Program (MEXT Q-LEAP) under Grant No. JPMXS0118069605.

\appendix

\section{Calculation of time averages\label{AppendixB}}
The purpose of this appendix is to provide a formal dervation of the time average of observables. This type of averages plays a very important role in describing equilibration in periodically-driven quantum system provided that the quasienergies show level repulsion.

\subsection{Formal derivation of time averages}
Let us consider the regime $\mathcal{M}\gg 1$ of time averages such as the one in \cref{eq:FloquetAvLocalMeanPhoton,eq:TwoPhotonProbabilityFloquet}. These averages contain quantities of the generic form
 \begin{align}
\label{eq:GenAverage}
\bar{Q}_{2N}=\frac{1}{\mathcal{M}}\sum^{\mathcal{M}-1}_{m=0}e^{-\mathrm{i}(\sum_{\zeta\in\boldsymbol{\zeta}}\xi_{\zeta}-\sum_{\eta\in\boldsymbol{\eta}}\xi_{\eta})mT/\hbar}
 \ ,
 \end{align}
 where $\boldsymbol{\zeta}=(\zeta_1,\zeta_2,\dots,\zeta_N)$ and $\boldsymbol{\eta}=(\eta_1,\eta_2,\dots,\eta_N)$. It is instructive to see the explicit expressions for $N=1$ and $N=2$
 \begin{align}
 \label{eq:TimeAV}
 \bar{Q}_2&=\frac{1}{\mathcal{M}}\sum^{\mathcal{M}-1}_{m=0}e^{-\mathrm{i}(\xi_{\alpha}-\xi_{\beta})mT/\hbar}
 \nonumber\\
 \bar{Q}_4&=\frac{1}{\mathcal{M}}\sum^{\mathcal{M}-1}_{m=0}e^{-\mathrm{i}(\xi_{\alpha}+\xi_{\lambda}-\xi_{\beta}-\xi_{\rho})mT/\hbar}
 \ .
 \end{align}
To calculate this type of averages in a formal way, it is useful to perform the sums explicitly. With this in mind, we consider a generic power series as follows
 \begin{align}
 \label{eq:GenTimeAV}
F(\phi-z)&=\frac{1}{\mathcal{M}}\sum^{\mathcal{M}-1}_{m=0}e^{-\mathrm{i}m (\phi-z)}
= \frac{1-e^{\mathrm{i}(z-\phi)\mathcal{M}}}{\mathcal{M}(1-e^{\mathrm{i}(z-\phi)})}
\nonumber\\
&=\left(\frac{1-e^{\mathrm{i}(z-\phi)\mathcal{M}}}{\mathcal{M}}\right)\sum_{l=0}^{\infty}\frac{\mathrm{i}^{l-1}B_l}{l!}(z-\phi)^{l-1}
 \ .
 \end{align}
where $\phi$ is a real variable and $z=z_{R}+\mathrm{i}z_{I}$ with $z_{I}\geq 0$ is a complex number added for convergence of the series. In this derivation, we also have used the generating function of the $B_l$ the Bernoulli numbers~\cite{gradshteyn2014table}
\begin{equation}
    \label{eq:Bernoulli}
         \frac{1}{1-e^{\mathrm{i}(z-\phi)}}= \sum_{l=0}^{\infty}\frac{\mathrm{i}^{l-1} B_l }{l!}(z-\phi)^{l-1}
\end{equation}
Next, let us carefully investigate the convergence of the power series in \cref{eq:GenTimeAV} in the complex plane. As a first step, we can notice that $F(0)=1$ when $z=\phi$. The second step is to look at the series far from the point $z=\phi$ defining a scale $R=|z-\phi|^{-1}$. Due to the exponential decay $e^{-z_{I}\mathcal{M}}$, the exponential term in \cref{eq:GenTimeAV} vanishes for $\mathcal{M}\gg R$ provided that $|z-\phi|\neq 0$. By using this, we can obtain the bound
\begin{equation}
    \label{eq:Bound}
         |F(1/R)|\leq \frac{1}{\mathcal{M}}\sum_{l=0}^{\infty}\frac{\mathrm{i}^{l-1}B_l}{l!}R^{1-l}
 \ . 
\end{equation}
This clearly tends to zero when $\mathcal{M}\gg R$. 
The trick to calculate the average is to take the limit $|z|\rightarrow 0$. This allows us to approach the resonance condition $\phi=0$ in the complex plane. After taking the limit $|z|\rightarrow 0$, we define a real function $F(\phi)$, from the previous discussion, we know that $F(\phi)=1$ if $\phi=0$ and $F(\phi)= 0$ when $\phi\neq 0$. Keeping this in mind, we obtain 
 \begin{align}
 \label{eq:FinalTimeAV}
 \bar{Q}_2&=\delta_{\alpha,\beta}
 \nonumber\\
 \bar{Q}_4&=\delta_{\alpha,\beta}\delta_{\lambda,\rho}+\delta_{\alpha,\rho}\delta_{\lambda,\beta}
 \ .
 \end{align}
The first expression follows from the generic Floquet spectrum~\cite{hamma2021a} by setting $\phi=(\xi_{\alpha}-\xi_{\beta})T/\hbar=T/\tau_{\alpha,\beta}$. From this equation, it is clear the meaning of the scale $R=\tau_{\alpha,\beta}/T$ that tells us how many periods we need to wait to obtain a well-defined time average, as we discussed in the main text. To obtain the second equation, we need to be careful because $\phi=(\xi_{\alpha}+\xi_{\lambda}-\xi_{\beta}-\xi_{\rho})T/\hbar$. Thus, we can have a resonance $\alpha=\beta$ and an effective scale $R=\tau_{\lambda,\rho}/T$. Alternatively, we can have $\lambda=\rho$ and  $R=\tau_{\alpha,\beta}/T$. For this reason, we need to take care of the possible combinations defining the time scales $\mathcal{M}\gg R$ to perform the time average.
The averages discussed here resemble the results for time-independent systems in Ref.~\cite{Lydzba2022generalized}. To calculate the general time average in \cref{eq:2NSpectralFormFactor}, one needs to consider all the possible pairs for which one reaches the resonance $\xi_{\zeta}-\xi_{\eta}$ for $\zeta\in \boldsymbol{\zeta}$ and $\eta\in \boldsymbol{\eta}$. For the general expression of this time average, we refer the reader to Appendix $C4$ of Ref.~\cite{hamma2021a}. 

\subsection{Time averages and Floquet generalized Gibbs states}
Let us explain in detail what is the nature of time averages of local observables. With this aim, we  investigate what happens with the dynamics of a single particle initialized in the state $\ket{\psi(0)}=\hat{a}^{\dagger}_i\ket{\boldsymbol{0}}$ in the chaotic regime. After $m$ periods of the circuit, the evolution of the particle is given by 
 \begin{equation}
 \label{eq:SIFloquetBosonicOperator}
 \ket{\psi(mT)}=\sum_{\alpha}e^{-\mathrm{i}\xi_{\alpha}mT/\hbar}c_{i,\alpha}\ket{\alpha}
   \ .
 \end{equation}

To investigate thermalization in quadratic Hamiltonians~\cite{Lydzba2022generalized}, it is useful to define the time average of a single-particle observable $\hat{O}$. One example of this is $\hat{O}=\hat{n}_l$. The time-averaged expectation value of such a single-particle observable over a total number $\mathcal{M}$ of periods of the drive is given by
 \begin{align}
 \label{eq:SIFloquetAvLocalMeanPhoton}
 \bar{O}&=\frac{1}{\mathcal{M}}\sum^{\mathcal{M}-1}_{m=0}\langle\hat{O}(mT)\rangle=
 \nonumber\\
 &=\frac{1}{\mathcal{M}}\sum^{\mathcal{M}-1}_{m=0}\sum_{\alpha,\beta}e^{-\mathrm{i}(\xi_{\alpha}-\xi_{\beta})mT/\hbar}
c_{i,\alpha}c^*_{i,\beta} \bra{\beta}\hat{O}\ket{\alpha}
 \ .
 \end{align}
From this expression, we clearly see the relation to $\bar{Q}_2$ in \cref{eq:GenAverage}. In the last section, we obtained the average $\bar{Q}_2=\delta_{\alpha, \beta}$, which give us the time average 
 \begin{align}
 \label{eq:SIDiagonalEnsemble}
 \bar{O}&=\sum_{\alpha}
|c_{i,\alpha}|^2 \bra{\alpha}\hat{O}\ket{\alpha}=\text{tr}(\hat{\rho}_{\text{GGE}}\hat{O})
 \ ,
 \end{align}
where 
\begin{align}
\label{eq:SIFloquetGibbs}
    \hat{\rho}_{\text{GGE}}=\sum_{\alpha}\ket{\alpha}\bra{\alpha}e^{-\Gamma_{\alpha}\xi_{\alpha}}/Z
    ,
\end{align}
is a density matrix representing the Floquet generalized Gibbs ensemble~\cite{Ishii2018} . Here $\Gamma_{\alpha}=1/k_B T_{\alpha}$ and $T_{\alpha}$ are effective temperatures determined by the conserved quantities, while $Z$ is a normalization constant~\cite{Ishii2018}. The direct implication of this is that due to level repulsion, the weights satisfy the relation $|c_{i,\alpha}|^2\approx e^{-\Gamma_{\alpha}\xi_{\alpha}}/Z$ that we used in the main text.

\section{Connection to the kicked rotor\label{AppendixA}}

In this section we show that a classical limit of \cref{eq:SingleParticle} at the single-particle level realizes a kicked rotor~\cite{Altland1996,Ammann1988,dArcy2001,Iomin2002,Manai2015,Akridas-Morel2019,santhanam2022quantum}, which is an archetypal model for chaos. Hence, we will show that the time evolution under Hamiltonian Eq. \cref{eq:SingleParticle} (see Figs.~3,4) is associated with the destruction of quasi-periodic orbits and a transition to the chaotic regime of a classical Hamiltonian.

\subsection{Semiclassic Hamiltonian in the absence of disorder \texorpdfstring{($W=0$)}{}}
 
It is useful to consider periodic boundaries $\hat{a}_{M+1}=\hat{a}_1$ and transform to the reciprocal lattice with the discrete Fourier transform
\begin{equation}
\hat{A}_{k} \equiv \frac{1}{\sqrt{M}} \sum_{j=1}^{M} e^{-i b k j} \hat{a}_{j},
\end{equation}
where $k$ labels the crystal momentum and we set the lattice constant $b = 1$. 
In this case, the variable $k$ is discrete and satisfy the condition $k=2s\pi/M$ with integer $s$. Importantly, for a finite chain, the values of $k$ are restricted to the first Brillouin zone $-\pi\leq k\leq \pi - 2\pi /M$.
Using the identity 
\begin{equation}
\label{eq:Completeness}
\frac{1}{M} \sum_j e^{-i(k_2 - k_1)j} = \delta_{k_1,k_2},
\end{equation}
where $\delta_{k_1,k_2}$ is the Kronecker delta function, \cref{eq:SingleParticle} becomes
\begin{align} \label{eq:ham-k}
\hat{H}(t) = & \frac{1}{M} \sum_{j, k_1,k_2} \frac{\hbar\phi_j}{T}
e^{-i (k_1 - k_2) j}
\hat{A}_{k_1}^{\dagger} \hat{A}_{k_2}^{} \nonumber \\
& + 2 \hbar J(t) \sum_k   \cos (k_1)  \hat{A}_{k}^{\dagger} \hat{A}_{k}
\ ,
\end{align}
where $J(t)=\frac{\theta}{T} \sum^{\infty}_{m=-\infty}\delta(t/T-m)$.
Since \cref{eq:SingleParticle} is quadratic we may restrict our analysis to the single-particle subspace. 

A general single-particle state in the reciprocal lattice basis is given by
\begin{equation}
| \psi(t) \rangle \equiv \sum_k \psi_k(t) \hat{A}_k^{\dagger} |0\rangle,
\end{equation}
where $\hat{A}_k^{} |0\rangle = 0$ defines the vacuum $|0\rangle$ and $\psi_k(t)$ is a complex coefficient. The time-dependent Schr\"odinger equation $i \hbar \partial_t |\psi(t) \rangle = \hat{H}(t) |\psi(t) \rangle$ is given by
\begin{align} \label{eq:schro-k}
\sum_k i \hbar \partial_t \psi_k(t) \hat{A}_k^{\dagger} |0\rangle
= & \frac{1}{M} \sum_{j,k,k_1} \frac{\hbar\phi_j}{T} e^{-i(k_1-k)j} \psi_{k_1}(t) \hat{A}_{k_1}^{\dagger} |0\rangle \nonumber \\
& + 2 \hbar J(t) \sum_k   \cos (k)  \psi_k(t) \hat{A}_k^{\dagger} |0\rangle, 
\end{align}
where we used the commutation relation $[A_{k_1}^{}, A_{k_2}^{\dagger}] = \delta_{k_1,k_2}$ for bosons. \cref{eq:schro-k} defines an equation of motion for the coefficients $\psi_k(t)$, given by
\begin{align} \label{eq:eigen-k}
\hbar \partial_t \psi_k(t) = \frac{1}{M}\sum_{j,k_1} \frac{\hbar\phi_j}{T} e^{-i(k_1 -k) j} \psi_{k_1}(t) 
+ 2 \hbar J(t) \cos (k)  \psi_k(t).
\end{align}

To obtain the semiclassical Hamiltonian, it is useful to calculate the following expression 
\begin{align}
         \label{eq:InfVolPosition}
         \sum_{j,k_2}\frac{je^{-\mathrm{i}(k_1-k_2) j}}{M^2}\psi_{k_2}(t) &=\lim_{\epsilon\rightarrow{0}}\frac{\mathrm{i}}{M^2}\sum_{j,k_2}\frac{e^{-\mathrm{i}(k_1-k_2) j}(e^{-\mathrm{i}\epsilon j}-1)}{\epsilon}\psi_{k_2}(t)
         \ .
\end{align}
To be mathematically precise, the right hand side of this equation can be interpreted as the derivative of the function
\begin{align}
         \label{eq:UsefulTrick}
          g(\epsilon)  &=\sum_{j,k_2}\frac{e^{-\mathrm{i}(k_1-k_2+\epsilon) j}}{M^2}\psi_{k_2}(t) = \sum_{k_2}\left[\frac{e^{\mathrm{i}\epsilon}-e^{-\mathrm{i}\epsilon M}}{e^{\mathrm{i}\epsilon}-e^{-\mathrm{i}(k_1-k_2) }}\right]\frac{\psi_{k_2}(t)}{M^2}
           \ 
\end{align}
with respect to $\epsilon$  and evaluated at $\epsilon=0$ as follows
\begin{align}
         \label{eq:UsefulTrick2}
         \mathrm{i} \frac{d g(\epsilon)}{d\epsilon}\Bigr|_{\epsilon=0}  &=\sum_{j,k_2}\frac{je^{-\mathrm{i}(k_1-k_2) j}}{M^2}\psi_{k_2}(t) 
          \ .
\end{align}

So far all the calculations are exact. But now we will carefully look at \cref{eq:InfVolPosition,eq:UsefulTrick} in the large-volume limit $M\gg1$ and make some approximations. As a first step, approximate the discrete sums in \cref{eq:UsefulTrick} using an integral in the limit $M\gg1$ as follow
\begin{align}
         \label{eq:UsefulIdentityDiscreteDelta}
        g(\epsilon)   &=\frac{1}{M}\sum_{k_2}\frac{1}{M \bar{\Delta}_k}\left[\frac{e^{\mathrm{i}\epsilon}-e^{-\mathrm{i}\epsilon M}}{e^{\mathrm{i}\epsilon}-e^{-\mathrm{i}(k_1-k_2) }}\right]\psi_{k_2}(t)\bar{\Delta}_k
          \nonumber \\
          &\approx\frac{1}{M}\int^{\pi}_{-\pi}\delta(k_1-k_2+\epsilon)\psi(k_2,t) dk_2  =\frac{\psi(k_1+\epsilon,t)}{M}              \ ,
\end{align}
where we defined $\psi(k,t)$ as the continuous version of $\psi_{k}(t)$. We also considered the volume element in the reciprocal space $\Delta_k=2\pi/M$ and approximate the integrand Kernel by using a Dirac delta $\delta(k_1-k_2+\epsilon)$ in the limit $M\gg1$ when $\Delta_k\rightarrow 0$. In the last derivation, we approximated the discrete sums using an integral in the limit $M\gg1$ as follows
\begin{align}
         \label{eq:DiscreteApprox}
\sum_k f(k)=\frac{1}{\bar{\Delta}_k}\sum_k  f(k)\bar{\Delta}_k\approx \frac{M}{2\pi} \int^{\pi}_{-\pi}f(k)dk
           \ .
\end{align}
This procedure allows us to approximate the sum in \cref{eq:InfVolPosition} in terms of a scaled derivative of the wave function $\psi(k_1,t)$ as follows
\begin{align}
         \label{eq:InfVolPositionImproved}
         \sum_{j,k_2}\frac{je^{-\mathrm{i}(k_1-k_2) j}}{M^2}\psi_{k_2}(t) 
          &\approx \lim_{\epsilon\rightarrow{0}} \frac{\mathrm{i}}{M}\frac{\psi(k_1+\epsilon,t)-\psi(k_1,t)}{\epsilon}
          \nonumber \\
          &\approx\frac{\mathrm{i}\partial_{k_1}\psi(k_1,t)}{M}
            \ ,
\end{align}
where used $\psi(k_1+\epsilon,t)\approx\psi(k_1,t)+\epsilon\partial_{k_1}\psi(k_1,t)$.

Hence, by using \cref{eq:InfVolPositionImproved} we can approximate the sum in \cref{eq:eigen-k} as follows

\begin{equation}
 \frac{1}{M}\sum_{j,k1} \frac{\hbar\phi_j}{T} e^{-i(k_1-k)j} \psi_{k_1}(t) \approx \frac{4 \hbar \Phi}{T} \left( \frac{i\partial_k}{M} - \frac{1}{2}  \right)^2 \psi_k(t).
\end{equation}

Defining the position as $\hat{q} \equiv b \hat{x} \equiv i\partial_k / M$ and momentum as $\hat{p} \equiv \hbar k / b$ (keeping in mind that $b=1$), we recover the canonical commutation relation $[\hat{q}, \hat{p}] = i \hbar_{\mathrm{eff}}$, with effective Planck constant $\hbar_{\mathrm{eff}} \equiv \hbar / M$. Hence, for $M \gg 1$, position and momentum behave like commuting classical variables and may be replaced with real numbers $\hat{x} \rightarrow x$. In this limit we obtain the classical Hamiltonian
\begin{align} \label{eq:ham-classical}
\mathcal{H}_{\mathrm{C}}(k,x,t) & = \frac{4 \hbar \Phi}{T} \left( x - \frac{1}{2} \right)^2 + 2 \hbar J(t)  \cos (k)  \nonumber \\
& =  \frac{4 \hbar \Phi}{T} \left( x - \frac{1}{2} \right)^2 +  \frac{2 \hbar\theta}{T}   \cos (k)  \sum^{\infty}_{m=-\infty} \delta(t/T - m),
\end{align}
where $(k,x)$ are classical canonical variables in phase space.

Next, let us calculate the classical equations of motion for the Hamiltonian given by \cref{eq:ham-classical}
 \begin{align}
 \label{eq:HamiltonEqs}
         \frac{\partial x}{\partial t}&= \frac{\partial \mathcal{H}(x,k,t)}{\hbar \partial k}=-\frac{2\theta}{T} \sin (k)\sum^{\infty}_{m=-\infty}\delta(t/T-m)
 \nonumber 
 \\
\hbar \frac{\partial k}{\partial t}&=-\frac{\partial \mathcal{H}(x,k,t)}{ \partial x}=-\frac{8 \hbar \Phi}{T}\left(x-\frac{1}{2}\right)
 \ .
 \end{align}
The integration of the equations of motion during a period $T$ of the drive gives us a discrete map that give us the dynamics at stroboscopic times
 \begin{align}
 \label{eq:ClassStrobDyn-original}
       x_{n+1}&= x_{n}-2\theta \sin (k_n)
 \nonumber 
 \\
k_{n+1}&=k_{n}-8\Phi\left(x_{n+1}-\frac{1}{2}\right)
 \ .
 \end{align}
Next, let us define the new coordinate $X_n=-8\Phi\left(x_{n}-\frac{1}{2}\right)$  in such a way that the equations of motion for this new variables are given by
 \begin{align}
 \label{eq:ClassStrobDyn}
      X_{n+1}&= X_{n}+\bar{K} \sin (k_n)
 \nonumber 
 \\
k_{n+1}&=k_{n}+X_{n+1}
 \ ,
 \end{align}
where $\bar{K}=16\theta\Phi$.

\subsection{Discussion about the relation to the Kicked rotor}

One can identify \cref{eq:ham-classical} with the model of a kicked rotor~\cite{Altland1996,Ammann1988,dArcy2001,Iomin2002,Manai2015,Akridas-Morel2019,santhanam2022quantum} with $x-1/2 \rightarrow p$ playing the role of the momentum and $k \rightarrow \theta$ the phase. \cref{eq:ham-classical} can be written as
\begin{equation} \label{eq:ham-kicked}
\mathcal{H}(\theta,p,t) = \frac{p^2}{2I} + K \cos\Theta \sum^{\infty}_{m=-\infty} \delta(t/T - n),
\end{equation}
where $I = T / (8\hbar \Phi)$ is the moment of inertia and $K = 2 \hbar \theta/T$ is the kicking strength. After integrating the equations of motions one obtains the discrete Chirikov map
\begin{align}
P_{n+1} = P_n + \bar{K} \sin \Theta_n \nonumber \\
\Theta_{n+1} = \Theta_n + P_{n+1},
\end{align}
where the angular momentum is $P \equiv pT / I$ and the renormalized kicking strength is $\bar{K} \equiv K T^2 / I=16\theta\Phi$ as defined previously.

The equations of motion of \cref{eq:ham-kicked} depend on a single parameter $\bar{K}$. It has the same structure as the kicked rotor conventionally found in textbooks, but the topology of the phase space differs. In a conventional kicked rotor the phase space lives on a torus because $P_n$ is taken modulo $2\pi$. In the kicked rotor defined by \cref{eq:ham-kicked}, the domain of $P_n$ is any real number and hence the topology of the phase space is a cylinder.

Hence, we can identify the classical limit of \cref{eq:SingleParticle} with the dynamics of the kicked rotor. When $\bar{K} = 0$ the dynamics is regular and the system shows only periodic orbits. The delta kicks proportional to $\bar{K}$ break the periodic orbits inducing a transition to chaos. For example, for $\bar{K}=4$, the system shows a mixed phase space with regular islands. The latter are completely absent when $\bar{K}=7$ and the system is fully chaotic. 

Some comments about the topology of the phase space and the definition of the coordinates $X_n$ and $k_n$ in \cref{eq:ClassStrobDyn} are in order. In the derivation of the semiclassical limit we have assumed periodic boundary conditions such that the position is in the domain $0\leq x_n<1$ and the momentum naturally is restricted to $-\pi<k_n<\pi$. Topologically, this defines a torus.  Consequently, the rescaled coordinates are defined in the domain  $-4\Phi \leq X_n<4\Phi $ and  $-\pi<k_n<\pi$.

\end{document}